\documentclass[twocolumn,a4paper,showpacs,preprintnumbers,amsmath,amssymb,pre]{revtex4-1}
\usepackage{graphicx}
\usepackage{dcolumn}
\usepackage{subfigure}
\usepackage{bm}
\usepackage{setspace}
\usepackage{fancyhdr}
\usepackage{amsmath}
\usepackage{color}
\usepackage{epstopdf}
\usepackage{psfrag}

\begin{document}

\title{Breakdown of the large-scale wind in $\Gamma=1/2$ rotating Rayleigh-B\'enard flow}
\author{Richard J.A.M. Stevens, Herman J.H. Clercx, Detlef Lohse}
\affiliation{
$^1$Department of Science and Technology and J.M. Burgers Center for Fluid Dynamics, University of Twente, P.O Box 217, 7500 AE Enschede, The Netherlands,\\
$^2$Department of Applied Mathematics, University of Twente, Enschede, The Netherlands\\
$^3$Department of Physics and J.M. Burgers Centre for Fluid Dynamics, Eindhoven University of Technology, P.O. Box 513, 5600 MB Eindhoven, The Netherlands
}
\date{\today}

\begin{abstract}
Experiments and simulations of rotating Rayleigh-B\'enard convection in cylindrical samples 
have revealed an increase in heat transport with increasing rotation rate. This heat transport 
enhancement is intimately related to a transition in the turbulent flow structure from a 
regime dominated by a large-scale circulation (LSC), consisting of a single convection roll, 
at no or weak rotation to a regime dominated by 
vertically-aligned vortices at strong rotation. 
For a sample with an aspect ratio $\Gamma = D/L = 1$ (D is the sample diameter and L its 
height) the transition between the two regimes is indicated by a strong decrease in the LSC 
strength. In contrast, for $\Gamma = 1/2$ Weiss and Ahlers 
[J. Fluid Mech. {\bf{688}}, 461 (2011)] revealed the presence of 
a LSC-like sidewall temperature signature beyond the critical rotation rate. They 
suggested that this might be due to the formation of a two-vortex state, in which one 
vortex extends vertically from the bottom into the sample interior and brings up warm fluid, 
while another vortex brings down cold fluid from the top; this flow field would yield a  sidewall temperature signature similar to that of the LSC. Here we show by direct numerical 
simulations for $\Gamma= 1/2$ and parameters that allow direct comparison with experiment 
that the spatial organization of the vertically-aligned vortical structures in the 
convection cell do indeed yield (for the time average) a sinusoidal variation of the 
temperature near the sidewall, as found in the experiment. This is also the essential 
and non-trivial difference with the $\Gamma=1$ sample, where the vertically-aligned 
vortices are distributed randomly.
\end{abstract}

\maketitle

\section{Introduction} \label{chapter1}
The flow of a fluid heated from below and cooled from above, better known as Rayleigh-B\'enard 
(RB) convection \cite{ahl09,loh10}, is the classical system to study heat transfer phenomena. 
For given aspect ratio $\Gamma\equiv D/L$ (D is the sample diameter and L its height) and 
given geometry, its dynamics is determined by the Rayleigh number 
Ra=$\beta g\Delta L^3 /(\kappa \nu)$ and the Prandtl number Pr=$\nu/\kappa$. Here, $\beta$ is the 
thermal expansion coefficient, g is the gravitational acceleration, $\Delta$ is the temperature 
difference between the horizontal plates, and  $\nu$ and $\kappa$ are the kinematic viscosity 
and thermal diffusivity, respectively. The response of the system is expressed by the 
dimensionless heat transfer, that is the Nusselt number Nu, and the Reynolds number Re. 
We will first give a brief introduction into the concept of the large scale circulation 
(LSC) (Section \ref{section1_1}) before we 
provide a brief summary of some important concepts and properties of rotating RB convection in 
Section \ref{section1_3}. Subsequently, in Section \ref{section1_4} we discuss the interesting aspects 
that are addressed in the present paper. 
\subsection{The LSC in Rayleigh-B\'enard convection} \label{section1_1}
\begin{figure*}
\centering
\subfigure[]{\includegraphics[width=0.31\textwidth]{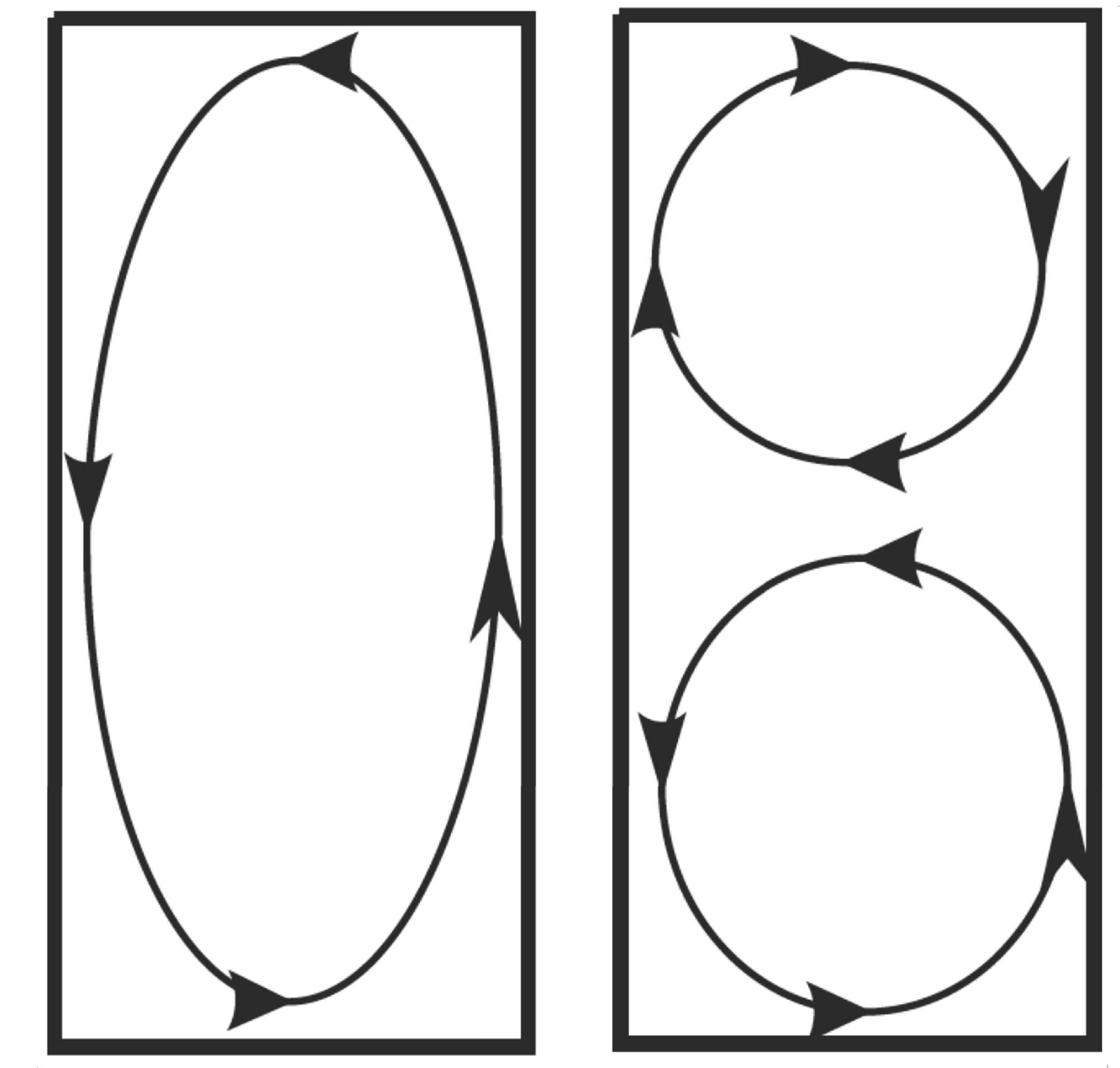}}
\subfigure[]{\includegraphics[width=0.31\textwidth]{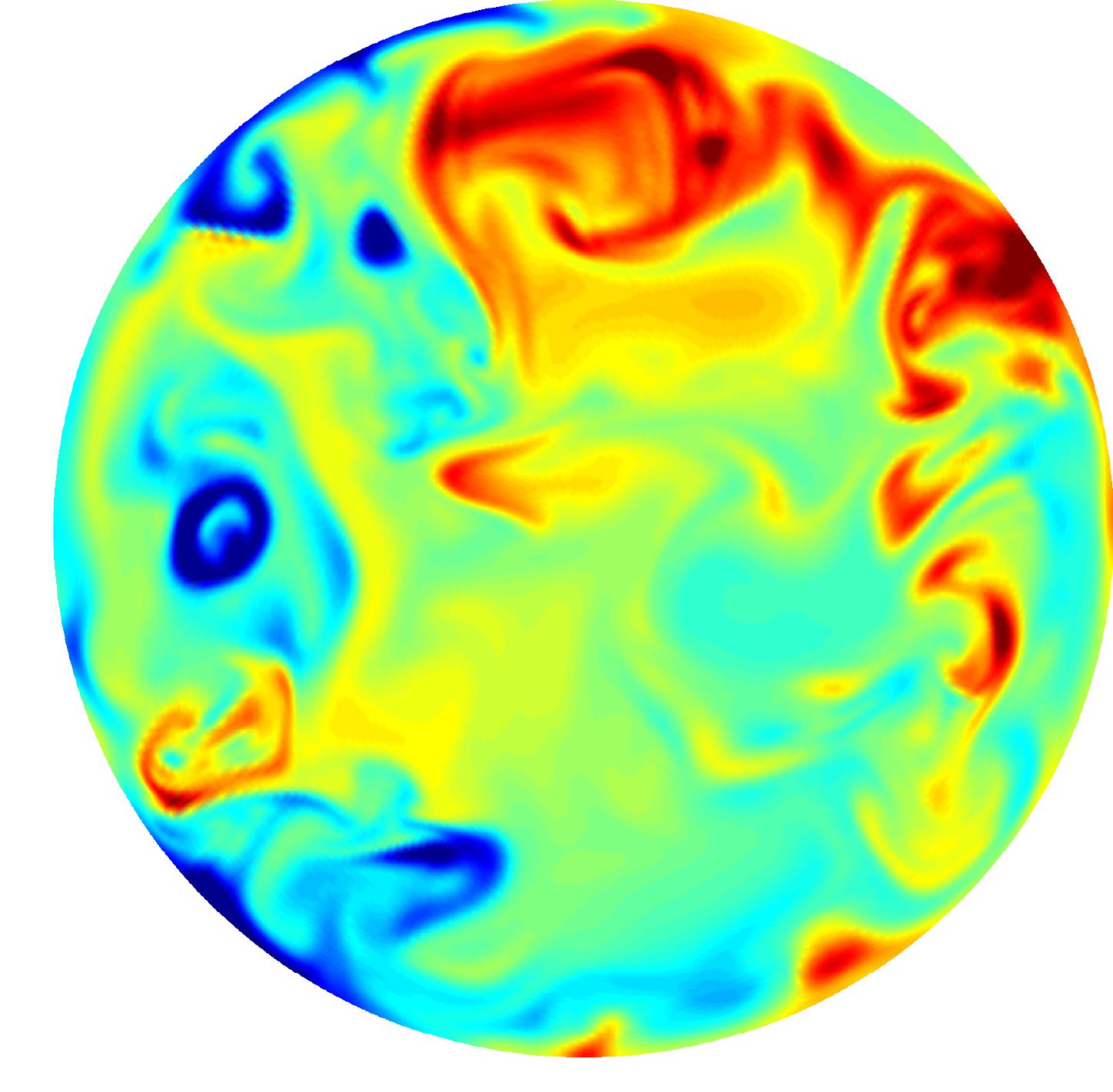}}
\subfigure[]{\includegraphics[width =0.31\textwidth]{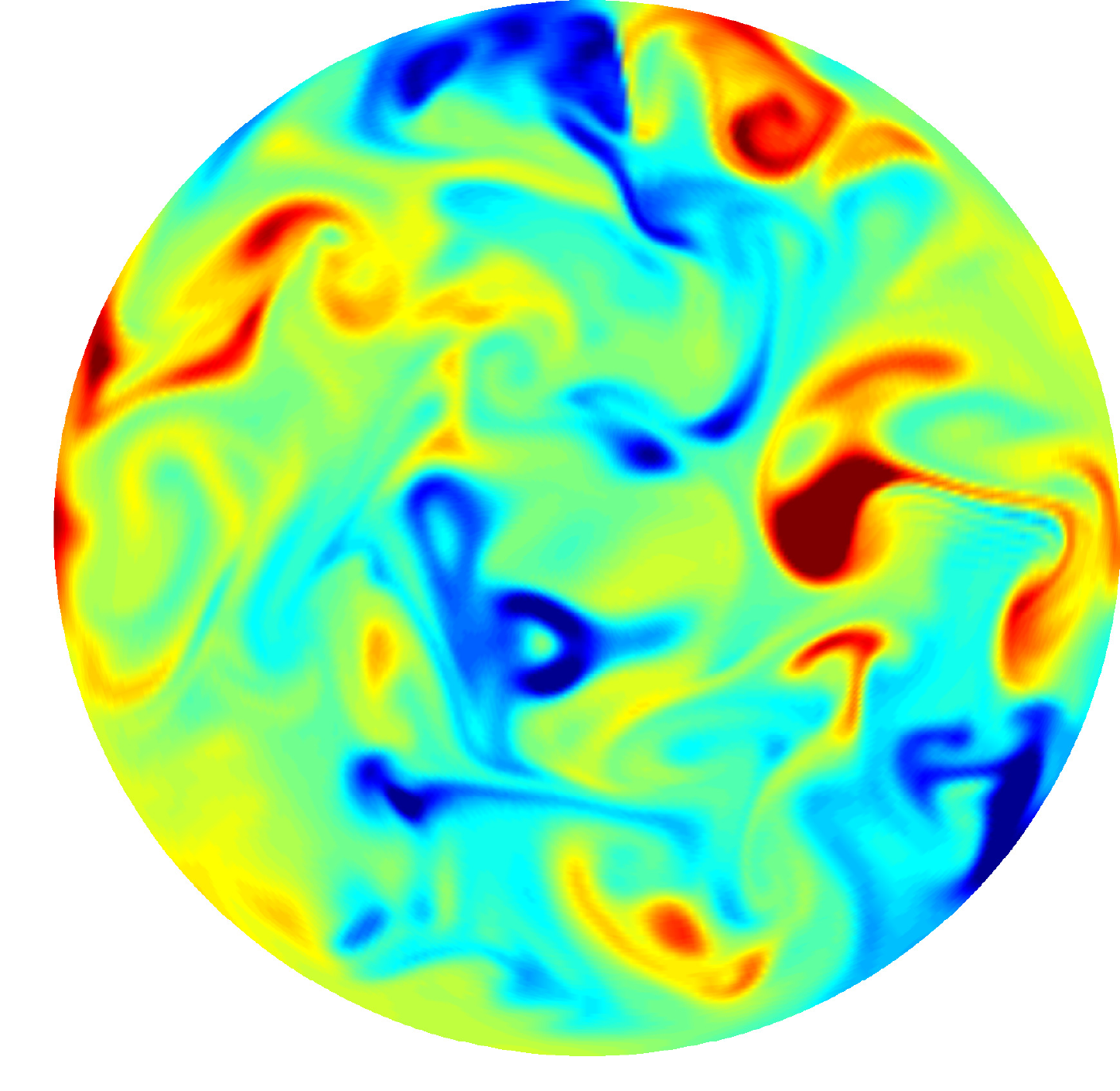}}
\caption{(Color online) a) Sketch of a single roll state (left) and a double roll state (right) of the large scale circulation in a $\Gamma=1/2$ sample. (b,c) Visualization of the  temperature field in the horizontal mid-plane of a cylindrical convection cell for  Ra=2.91 $\times 10^8$ and Pr=4.38 in a $\Gamma=1$ sample \cite{ste09}. The red and  blue areas indicate warm and cold fluid, respectively. (b) At a slow rotation rate  (1/Ro=0.278), and below the transition: The warm up-flow (red) and cold down-flow (blue) reveal the existence of a single convection roll superimposed upon turbulent fluctuations. (c) At a somewhat larger rotation rate (1/Ro=0.775), above the  transition: Here, the vertically-aligned vortices cause disorder on a smaller length scale in the horizontal plane.}
\label{figure1}
\end{figure*}

Due to the temperature difference between the horizontal plates warm plumes rise from the hot 
bottom plate and cold plumes sink from the cold top plate. In our cylindrical convection cell the 
collection of rising and sinking plumes organize as follows: the plumes with  warm fluid collect 
and flow up on one side of the cell and plumes with cold fluid flow down on the opposite side of 
the convection cell setting up a large-scale mean flow in the cell. This fluid motion is better 
known as the LSC, see the diagram in figure \ref{figure1}a. A recent 
review on the LSC, its properties and dynamics is provided in Ref.\ \cite{ahl09}.\\
Keeping the sketch of the LSC in mind we may expect hot rising fluid on 
one side and cold sinking fluid at the opposite side of the cell which may be visualized by 
horizontal temperature snapshots. A visualization of an instantaneous temperature field at mid 
height obtained in a simulation indeed gives an impression of the structure of the LSC, see 
figure \ref{figure1}b. In experiments the LSC is measured by thermistors in the sidewall, see for example Brown {\it et al.}\ \cite{bro05b}, or by small thermistors that are placed in the flow at 
different azimuthal positions and different heights, see for example Xia {\it et al.}\ \cite{xi09}. 
Since the LSC transports warm (cold) fluid from the bottom (top) plate up (down) the sidewall, 
the thermistors can detect the location of the up-flow (down-flow)  by showing a relatively high 
(low) temperature effectively resulting in a sinusoidal azimuthal temperature profile.\\

In order to analyze the sidewall temperature profile 
Stevens {\it et al.}\ \cite{ste10c} proposed a quantitative measure for the temperature signature, 
which they called the relative LSC strength in their study on the effect of plumes on measuring 
the LSC in samples with $\Gamma=1$ and $\Gamma=1/2$ in non-rotating RB convection. This measure is based on the energy 
in the different Fourier modes of the measured or computed azimuthal temperature profile at or 
nearby the sidewall, as
\begin{equation}\label{Eq Relative Strength LSC}
    \bar{S}_k = \mathrm{max}\left( \frac{\frac{N\langle E_1 \rangle}{\langle E_{tot}\rangle} - 1}{N-1}, 0 \right)~.
\end{equation}
The subscript $k$ indicates the height at which $\bar{S}_k$ is determined, that is for $k=b$ at 
$z/L=0.25$, for $k=m$ at $z/L=0.50$, and for $k=t$ at $z/L=0.75$. In equation 
(\ref{Eq Relative Strength LSC}) $\langle E_1\rangle \equiv\int_{t_b}^{t_e} E_1 dt$ indicates the 
sum of the energy in the first Fourier mode over the time interval $[t_b,t_e]$ (from the 
beginning of the statistically steady part of the simulation at $t=t_b$ to the end of the 
simulation at $t=t_e$), and $\langle E_{tot}\rangle$ is defined similarly with $E_{tot}$ being the total 
energy in all Fourier modes. Furthermore, $N$ is the total number of Fourier modes that can be 
determined (depending on the number of azimuthal probes). From its definition it follows that 
$0\le \bar{S}_k \le 1$. Concerning the limiting values: $\bar{S}_k=1$ means the 
presence of a pure azimuthal cosine profile and $\bar{S}_k=0$ indicates that the 
magnitude of the cosine mode is equal to (or weaker than) the value expected from a 
random noise signal, see Ref.\ \cite{ste10c}. Hence, a value for 
$\bar{S}_k$ of about $0.5$ or higher is a signature that a cosine fit on average 
is a reasonable approximation of the data set, as then most energy in the signal 
resides in the first Fourier mode. In contrast, $\bar{S}_k$ well below $0.5$  indicates 
that most energy resides in the higher Fourier modes. In section \ref{section1_3} 
we discuss how this method is used to determine the existence of different turbulent 
states in rotating RB convection.
\subsection{Rotating Rayleigh-B\'enard convection} \label{section1_3}
The case where the RB system is rotated around its vertical axis at an angular speed 
$\Omega$ is used to study the effect of rotation on heat 
transport and flow structuring. Here we define the Rossby number as $Ro=U/(2\Omega L)$, with $U = \sqrt{\beta g \Delta L}$ being the free-fall velocity (mean bulk velocity) to make sure that Ro is the relevant parameter that determines when the formation of large scale vortices sets in. In the remainder of the paper we will indicate the dimensionless rotation rate by 1/Ro as this parameter increases with increasing rotation rate.

Since the experiments by Rossby \cite{ros69} it is known that rotation can enhance heat 
transport. Rossby measured an increase in the heat transport with respect to the 
non-rotating case of about $10\%$ when water is used as the convective fluid. This 
increase is counterintuitive as the stability analysis of Chandrasekhar \cite{cha81} has shown 
that rotation delays the onset to convection and from this analysis one would expect that 
the heat transport decreases. The mechanism responsible for this heat transport 
enhancement is Ekman pumping \cite{ros69,jul96b,vor02,kun08b,kin09,zho09b}, that is, 
due to rotation, rising or falling plumes of hot or cold fluid are stretched into 
vertically-aligned vortices that suck fluid out of the thermal boundary layers adjacent to 
the bottom and top horizontal plates. Evidence for the existence of vertically-aligned 
vortices was reported 15-20 years ago \cite{bou90,zho93,sak97}. Sakai was the 
first who confirmed with flow visualization experiments that there is a typical ordering of 
vertically-aligned vortices under the influence of rotation. For higher rotation rates a strong heat transport reduction, 
due to the suppression of the vertical velocity fluctuations by rotation, is found. 
After Rossby \cite{ros69} many experiments have confirmed this 
general picture qualitatively and, in recent years, also quantitatively, see for example Refs.\ 
\cite{zho93,jul96,liu97,vor02,kun08b,kin09,liu09,zho09b,ste09,zho10c,wei10,wei11,wei11b,kun11}.\\
Next to experiments there have been a number of numerical studies of rotating RB convection 
\cite{kun08b,kun10,kun10b,kun11,zho09b,wei10,ste09,ste10a,ste10b,ste11b}. These studies focussed 
on the influence of rotation on heat transport and the corresponding changes in the flow 
structure. Zhong {\it et al.}\ \cite{zho09b} and Stevens {\it et al.}\ \cite{ste09,ste10a} used 
results from experiments and simulations in a $\Gamma=1$ sample to study the influence of Ra 
and Pr on the effect of Ekman pumping, and thus on heat transport.\\ 
For an overview of the parameter regimes that are covered in simulations and 
experiments during the recent decades we refer to the figures 1 of Stevens 
{\it et al.}\ \cite{ste10b,ste11b}. 
These phase diagrams reveal that there are two approaches to analyze rotating RB convection. The 
first approach is to vary the rotation rate while the temperature difference between the two 
horizontal plates is fixed. This means that one varies 1/Ro, while Ra is kept constant. The 
second approach is to keep the ratio between the viscous force to the Coriolis force, indicated 
by the Ekman (Ek) or Taylor (Ta) number, fixed. 
In almost all simulations and experiments the Pr number is kept constant and a fixed Ek number 
then means that Ra and 1/Ro are varied. This approach is followed by several 
authors \cite{liu97,liu09,liu11,kin09,kin12,sch09,sch10}. Here we take 
the first approach, that is,  vary the rotation rate while the temperature difference between the 
plates is fixed.

When heat transport enhancement as function of the rotation rate is considered, a 
typical division in three regimes is observed. Namely regime 
I (weak rotation), where no heat transport enhancement is observed, regime II (moderate 
rotation), where a strong heat transport enhancement is found, and regime III (strong 
rotation), where the heat transport starts to decrease \cite{bou90,kun10,kun11}. Based 
on an experimental and numerical study on the properties of the LSC in rotating RB in a 
$\Gamma=1$ sample Kunnen {\it et al.}\ \cite{kun08b} showed that there is no LSC in 
regimes II and III. Subsequently, Stevens {\it et al.}\ \cite{ste09} showed for a similar setting 
that the heat transport enhancement at the start of regime II sets in as a sharp 
transition (at a critical value of the inverse Rossby number, 1/Ro$_c$). 
They showed that in experiments the transition is indicated by changes in the 
time-averaged LSC amplitudes, i.e. the average amplitude of the cosine fit to the 
azimuthal temperature profile at the sidewall, and the vertical temperature gradient at 
the sidewall. We note that the formation of a mean vertical temperature gradient 
in rotating RB convection was already observed earlier, see Refs. \cite{jul96,har99}. 
Later experiments and simulations \cite{kun11,wei11b} in a 
$\Gamma=1$ sample revealed that $\bar{S}_k$ is close to one before the onset of heat 
transport enhancement. In this regime the LSC is the dominant 
flow structure. After the onset vertically-aligned vortices become the dominant flow 
structure and then $\bar{S}_k$ quickly decreases to zero. The reason is the 
formation of many randomly positioned vertically-aligned vortices, and due to their 
random locations, see figure \ref{figure1}c, the cosine mode in the azimuthal temperature 
profile disappears.

In addition, simulation results have shown that the transition between the two different 
states is visible not only in the Nu number and the LSC statistics, but also by a 
strong increase in the number of vortices at the thermal boundary layer height 
\cite{ste09,ste11b,wei10,kun10b}. 
In addition, rotation changes the character of the kinetic boundary layer from a 
Prandtl-Blasius boundary layer at no or weak rotation to an Ekman boundary layer after 
the onset, which is revealed by the ${1/Ro}^{-1/2}$ scaling of the kinetic boundary layer 
thickness after the onset for $\Gamma=1$. Furthermore, Stevens {\it et al.}\ \cite{ste09} 
observed for this case an increase in the vertical 
velocity fluctuations at the edge of the thermal boundary layer, due to the effect of 
Ekman pumping, while a decrease in the volume averaged vertical velocity fluctuations 
is found due to the destruction of the LSC. 

\subsection{Issues addressed in the present paper}  \label{section1_4}
Most findings described above are based on experiments and simulations in $\Gamma=1$ 
samples. In the present paper we study rotating RB convection in an aspect ratio 
$\Gamma=1/2$ sample as experiments have revealed important differences with respect 
to the $\Gamma=1$ case. Recently, Weiss and Ahlers \cite{wei10b} revealed that in such a $\Gamma=1/2$ 
sample with water (Pr=4.38) at a range of Rayleigh numbers 
($2.3 \times 10^9 \leq Ra \leq 7.2 \times 10^{10}$) the flow can be either in a single 
roll (see left panel of figure \ref{figure1}a) or in a double roll state (see right panel of figure \ref{figure1}a) when no rotation is applied. Here, modest rotation 
(1/Ro$\lesssim$ 1/Ro$_c$), stabilizes the single roll 
state and suppresses the double roll state. Surprisingly, computation of $\bar{S}_k$  
reveals that after the onset of heat transport enhancement at 1/Ro$_c$ the cosine-signature of the temperature 
profile at the sidewall does not disappear. This may suggest that in a $\Gamma=1/2$ sample the 
single roll state continues to exist in the rotating regime, whereas it has been shown to 
disappear in a $\Gamma=1$ sample \cite{kun08b,kun11,ste09,wei11b}. As discussed by Weiss and 
Ahlers \cite{wei11,wei11b} the value of $\bar{S}_k\gtrsim 0.5$ does not allow us to 
distinguish between a single roll state and a two-vortex state, in which one vortex 
extends vertically from the bottom into the sample interior and brings up warm fluid, 
while another vortex transports cold fluid downwards at the opposite part of the 
cylindrical sample. Such a two-vortex state presumably 
results in a periodic azimuthal temperature variation close to the sidewall which cannot 
be distinguished from the temperature signature of a convection roll with up-flow and 
down-flow near the side wall. It is worthwhile to emphasize the distinction with the 
$\Gamma=1$ case with 1/Ro$>$1/Ro$_c$. Here, the vertically-aligned vortices 
emerging at the bottom plate (transporting hot fluid upwards) are distributed more or 
less randomly over the  horizontal cross section and similarly for the vortex tubes 
emanating from the top plate.
 
 \begin{table*}
  \centering
  \caption{Details for the simulations performed at Ra=$4.52\times 10^9$ and Pr=4.38 in an aspect ratio $\Gamma=1/2$ sample. The columns from left to right indicate the Rossby number 1/Ro, the Nusselt number ($Nu_f$) obtained after averaging the results of the three methods (see text) using the whole simulation length, and the Nusselt number ($Nu_h$) after averaging the results of the three methods over the last half of the simulation. The following column indicates the normalized temperature gradient at the sidewall $\Delta_w/\Delta$. The next three columns give the Nusselt number derived from the volume averaged value of the pseudo kinetic dissipation rate $\langle\epsilon_u^" \rangle$, and the kinetic $\langle\epsilon_u \rangle$ and thermal $\langle\epsilon_\theta \rangle$ dissipation rates compared to $Nu_f$ indicated in column two. The last column indicates the simulation length in dimensionless time units that is considered for the data collection; before this period each simulation has been run for $80-100$ time units in order to prevent transient effects.}\label{Table1}
\begin{tabular}{|c|c|c|c|c|c|c|c|}
\hline
1/Ro 	& $Nu_f$		& $Nu_h$ & $\Delta_w/\Delta$ & $\frac{\frac{\langle \epsilon_u^" \rangle}{\nu^3Ra/(Pr^2L^4)}+1}{Nu_f}$ 	&$\frac{\frac{\langle \epsilon_u \rangle}{\nu^3Ra/(Pr^2L^4)}+1}{Nu_f}$ & $\frac{\frac{\langle \epsilon_\theta \rangle}{\kappa\Delta^2/L^2}}{Nu_f}$ & $\tau_f$ \\
  \hline
$0$ 		& 103.82		& 104.03	&	-0.0848	&  	0.996	&	0.995	&	0.976 	& 330\\
$0.2778$ 	& 104.27		& 103.80	&	-0.0739	&    	0.985	&	0.976	&	0.974	& 330\\
$0.3571$ 	& 104.89		& 104.86	&	-0.0642	&	0.993	&	1.019	&	0.975	& 330\\
$0.4651$ 	& 105.58		& 105.45	&	-0.0645	&	0.985	&	0.976	&	0.973	& 330\\
$0.5263$ 	& 105.61		& 105.75	&	-0.0538	&	0.988	&	1.001	&	0.974	& 330\\
$0.5988$ 	& 105.86		& 106.09	&	-0.0474	&	1.001	&	1.000	&	0.976	& 330\\
$0.6897$ 	& 105.56		& 105.41	&	-0.0529	&	0.992	&	0.995	&	0.975	& 330\\
$0.7752$ 	& 106.14		& 105.58	&	-0.0469	&	1.002	&	0.984	&	0.976	& 330\\ 
$0.8696$ 	& 106.66		& 107.18	&	-0.0420	&	1.001	&	1.014	&	0.976	& 306\\
$1.0000$ 	& 106.22		& 106.03	&	-0.0547	&	0.983	&	0.971	&	0.974	& 252\\
$1.2500$ 	& 107.54		& 107.55	&	-0.0620	&	0.999	&	0.980	&	0.974	& 268\\
$1.5385$ 	& 108.49		& 108.22	&	-0.0811	&	0.980	&	0.990	&	0.973	& 282\\
$2.2222$ 	& 111.25		& 110.91	&	-0.0862	&	0.986	&	0.984	&	0.973	& 276\\
$3.3333$ 	& 111.93		& 112.19	&	-0.0946	&	1.002	&	0.989	&	0.974	& 311\\
$5.0000$ 	& 112.23		& 111.98	&	-0.1148 	&	0.991	&	0.986	&	0.971	& 330\\
$8.3333$ 	& 111.09		& 111.38	&	-0.1272 	&	0.986	&	0.970	&    	0.969	& 330\\
$10.000$ 	& 104.97		& 105.19	&	-0.1873 	&	1.004	&	1.047	&   	0.969	& 487\\
$12.500$ 	& 101.41		& 102.88	&	-0.1778	&	1.006	&	0.982	&    	0.971	& 330\\
  \hline
\end{tabular}
\end{table*}
\noindent From a retrospective point of view we can now conclude that Stevens {\it et al.}\ 
\cite{ste11b} already observed some first hints for the presence of a two-vortex state (see figure \ref{figure6} for an impression of a two-vortex state) in a $\Gamma= 1/2$ sample at 
Ra = $2.91 \times 10^8$, 1/Ro = 3.33 and  Pr = 4.38 (see figure 5a of that paper). 
However, an open question is whether this state prevails for a considerable time and for different values of 
Ra, Pr, and a range of rotation rates (in particular matching the settings of the experiments by Weiss and Ahlers 
\cite{wei11,wei11b}). Additionally, can we complement the indirect evidence for a 
two-vortex state as put forward by Weiss and Ahlers \cite{wei11,wei11b} with direct 
observations of the flow structures in the bulk flow? Can we also distinguish between a 
genuine two-vortex state and a flow structuring consisting of a few vertically-aligned 
vortices with hot rising fluid on one side of the cylindrical sample and a few of such 
vortices bringing cold fluid from the top plate downwards close to the opposite sidewall? 
In the present paper we report on direct numerical simulation results for 
$\Gamma = 1/2$, Ra = $4.52 \times 10^9$, and Pr = 4.38 over the range 
0 $\lesssim$ 1/Ro $\lesssim$ 12, corresponding to a subset of the Weiss-Ahlers experiments thus allowing direct comparison. 
The results show that a complex vortex state, which in the time average has the 
signature of a two-vortex state, persists over the range 
1/Ro$_c$ $\lesssim$ 1/Ro $\lesssim$ 12. This confirms that the observation of a sinusoidal azimuthal 
temperature profile near the sidewall can indeed be explained by the presence of Ekman 
vortices. In addition, to this general 
result, we report detailed data that permit a direct comparison of $S_t$, $S_m$, and $S_b$ 
(see figure \ref{figure4}) and the LSC amplitude (see figure \ref{figure5}) with 
experiment over a wide 1/Ro range. Furthermore, we also computed the Nusselt number 
and the vertical temperature gradient at the sidewall (see figure \ref{figure3}). 
All of these properties show excellent agreement between our simulations and the available 
experimental measurements; our simulations can therefore be used for exploration 
of the bulk and boundary layer flow structure which are not accessible in the current experiments.

We first discuss the numerical method in Section \ref{chapter2}, before 
we compare the simulation data with the experiments in Section \ref{chapter3}. In Section \ref{chapter4} we discuss flow diagnostics that can only be obtained in 
simulations and show that in a $\Gamma=1/2$ sample the vertically-aligned vortices 
arrange such that a sinusoidal azimuthal temperature profile close to the sidewall 
is formed. 

\section{Numerical method} \label{chapter2}
We simulate rotating RB convection for Ra=$4.52\times10^9$ and Pr=4.38 in an aspect 
ratio $\Gamma=1/2$ sample by solving the three-dimensional Navier-Stokes equations within 
the Boussinesq approximation. A constant temperature boundary condition is applied at the horizontal plates, while the sidewall is modeled as adiabatic. This case has been chosen to be as close as possible to the experiments performed by Weiss and Ahlers \cite{wei11,wei11b}.
General details about the numerical procedure can be found in Refs.\ \cite{ver96,ver99} and 
specific details concerning the (non)rotating RB simulations in Refs.\ \cite{zho09b,ste10a}.

In order to eliminate the effect of transients we discarded the information of the first 
$80-100$ dimensionless time units and the simulation lengths we mention refer to the 
length of the actual simulation, thus the period after this initialization. In Table 
\ref{Table1} we compare the Nusselt number averaged over the whole simulation length 
(denoted by $Nu_f$; the Nusselt number is based on the averages of three methods, i.e. 
the volume average of 
$Nu = (\langle u_z \theta \rangle_A -\kappa \partial_3 \langle \theta \rangle_A$)/ $\kappa \Delta L^{-1}$  
and the averages based on the temperature gradients at the bottom and top plate) with the 
Nusselt number averaged over half the simulation length ($Nu_h$).  For all cases these 
values $Nu_f$ and $Nu_h$ are converged within $1 \%$.\\
The simulations have been performed on a grid with $641\times 161 \times 641$ nodes in 
the azimuthal, radial, and axial direction, respectively. The grid allows for a very 
good resolution of the small scales both inside the bulk of turbulence and in the 
boundary layers where the grid-point density has been enhanced. We checked this by 
calculating the Nusselt from the volume-averaged kinetic energy dissipation rate 
$\langle \epsilon_u \rangle=\nu^3(Nu-1)Ra/(Pr^2L^4)$, and thermal dissipation rate 
$\langle \epsilon_\theta \rangle= \kappa\Delta^2 Nu /L^2$ as is proposed by Stevens 
{\it et al.}\ \cite{ste10}. In addition, we now also compare with the volume averaged 
value of $\epsilon_u^":= \bf{u} \cdot \nabla^2 \bf{u}$. The Nusselt number calculated 
from these quantities is always within a $5\%$ margin, and even much closer for most 
simulations, of $Nu_f$, which according to Stevens {\it et al.}\ \cite{ste10} indicates 
that the simulations are well-resolved, see Table \ref{Table1} for details. 
As argued by Shishkina {\it et al.}\ \cite{shi10} it is especially important to properly 
resolve the boundary layers. Our grid-point resolution in the boundary layers also satisfy 
their criteria for the rotating case (where kinetic 
boundary layers tend to become thinner with increasing rotation rate).

\section{Comparison with experiments} \label{chapter3}
\begin{figure}
  \centering
  \subfigure{\includegraphics[width=0.48\textwidth]{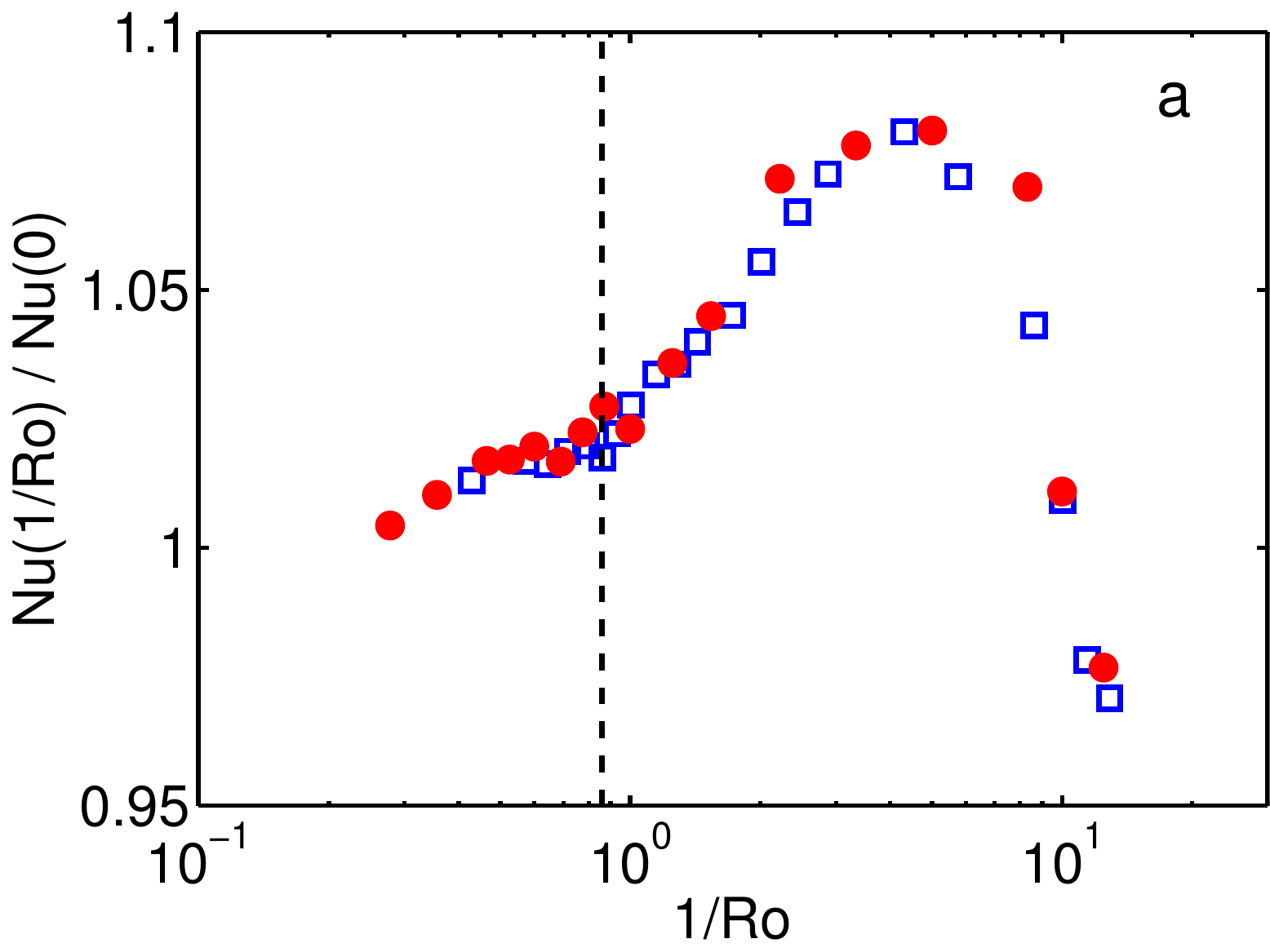}}
  \subfigure{\includegraphics[width=0.48\textwidth]{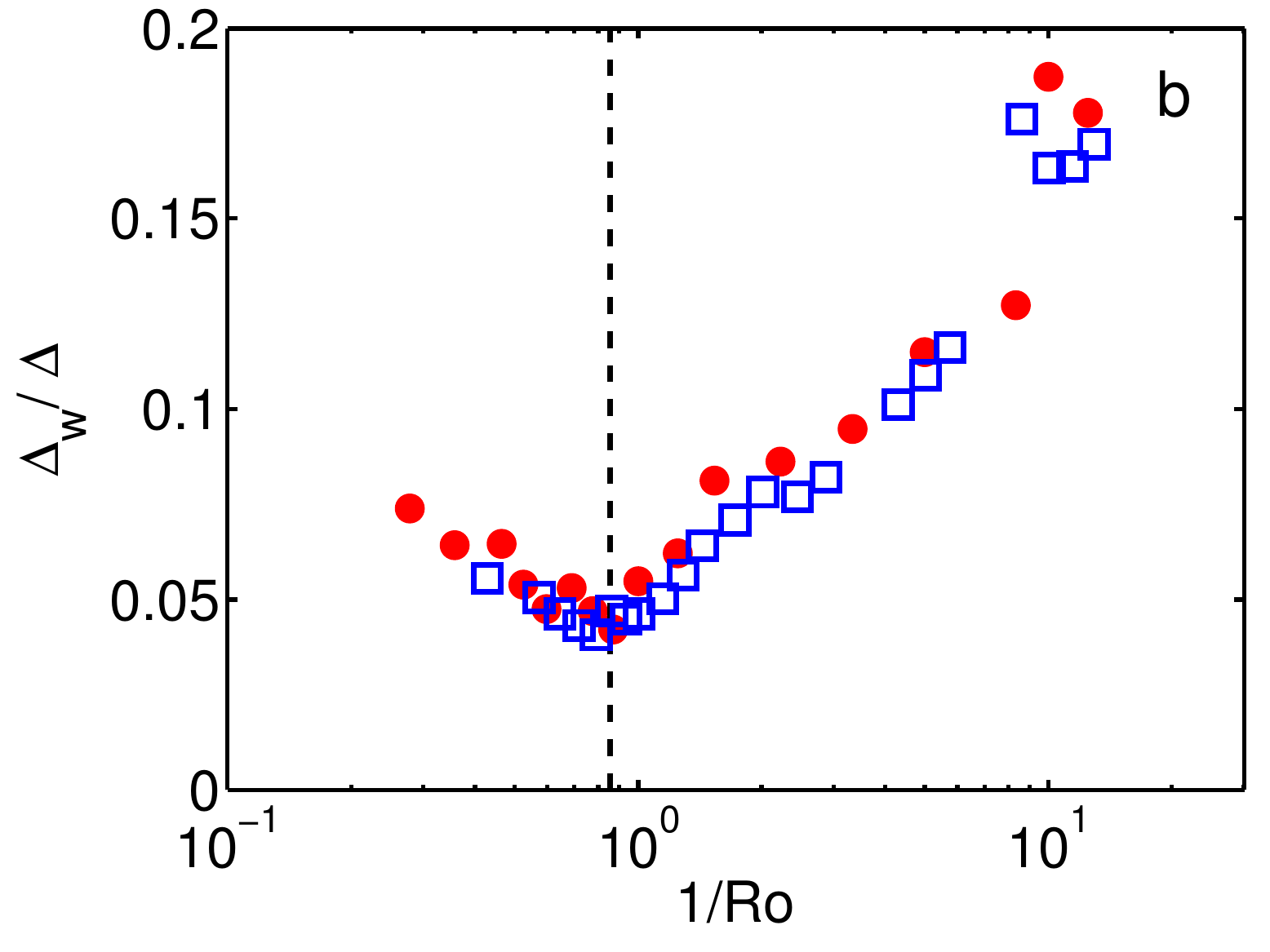}}
  \caption{(color online) (a) The heat transfer enhancement with respect to the non-rotating case, 
Nu(1/Ro)/Nu(0), and (b) the temperature gradient at the sidewall as function of the 
rotation rate 1/Ro for Ra=$4.52\times10^9$ and Pr=4.38 in a  $\Gamma=1/2$ sample. 
The experimental data from Weiss and Ahlers \cite{wei11b} are indicated by the blue open 
squares and the simulation results with the red solid circles. The vertical dashed lines 
indicates the position of the transition (at 1/Ro$_c$ $\approx$ 0.86)~\cite{wei10}.}
  \label{figure3}
\end{figure}

In figure \ref{figure3}a we show that the Nusselt number as function of the rotation 
rate Nu(1/Ro), with respect to the non-rotating value Nu(0), obtained in the 
simulations agrees excellently with the experimental result of Weiss and Ahlers \cite{wei11b}. In 
addition, we note that the absolute values differ less than $1\%$, which can be 
considered as an excellent agreement. The figure shows that a strong heat transport 
enhancement due to Ekman pumping sets in at a critical dimensionless rotation rate 
1/Ro$_c$ $\approx$ 0.86. 
For strong rotation rates, i.e. high values of 1/Ro, the expected decrease in the heat 
transport is observed. 
In figure \ref{figure3}b, we find that the normalized vertical temperature gradient at the 
sidewall at $z/L=0.50$, denoted by $\Delta_w/\Delta$, and calculated from the azimuthally 
and time averaged temperatures at the sidewall at $z/L=0.25$ and $z/L=0.75$, is also in 
good agreement with the measurements by Weiss and Ahlers \cite{wei11b}.

\begin{figure*}
  \centering
  \subfigure{\includegraphics[width=0.32\textwidth]{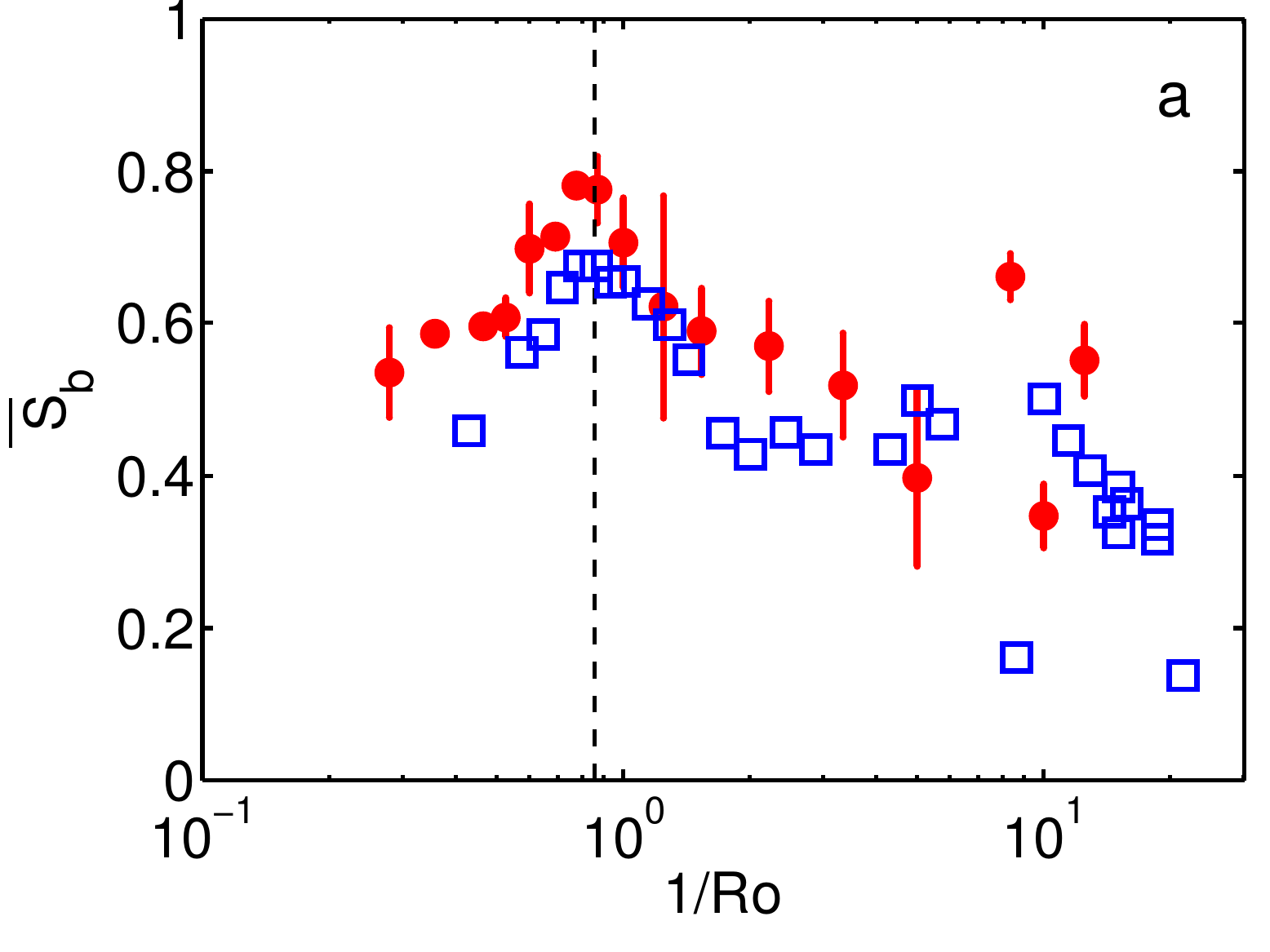}}
  \subfigure{\includegraphics[width=0.32\textwidth]{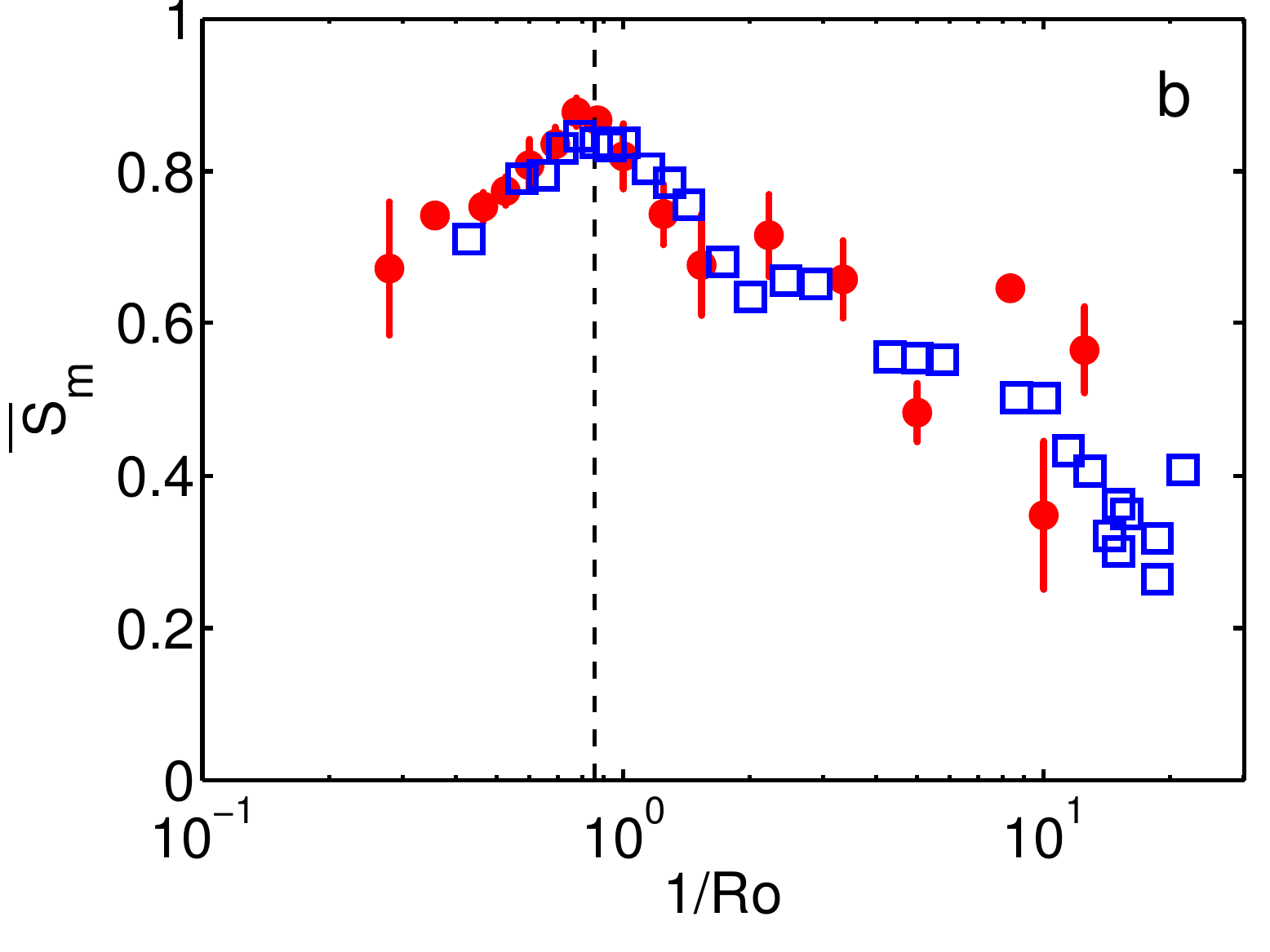}}
  \subfigure{\includegraphics[width=0.32\textwidth]{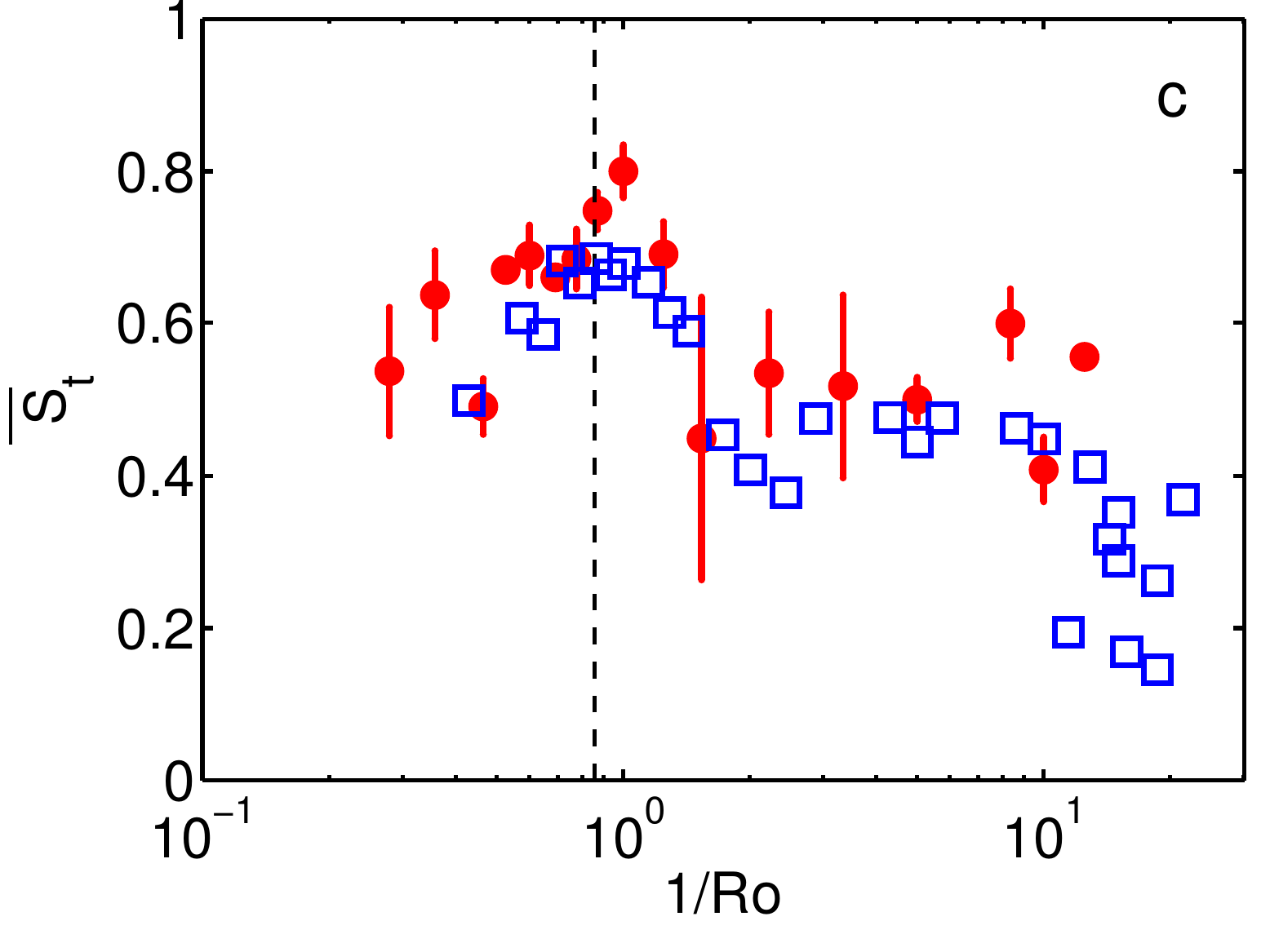}}  
  \caption{(color online) The values of (a) $\bar{S}_b$ (at $0.25z/L$), (b) $\bar{S}_m$ (at $0.50z/L$), and (c) 
$\bar{S}_t$ (at $0.75z/L$) as function of 1/Ro. The experimental data from Weiss and 
Ahlers \cite{wei11b} are indicated by blue open squares and the simulation results with red 
solid circles. The experimental and simulation results are for Ra=$4.52\times10^9$ and Pr=4.38 
in a $\Gamma=1/2$ sample. For the simulation data $\bar{S}$ is based on the temperature 
measurements of $64$ azimuthally equally spaced probes outside the sidewall boundary 
layer and for the experimental data it is based on temperature measurements of $8$ 
probes embedded in the sidewall. The vertical dashed line indicates the position of 
1/Ro$_c$ $\approx$ 0.86~\cite{wei10}.}
  \label{figure4}
\end{figure*}

\begin{figure}
  \centering
  \subfigure{\includegraphics[width=0.48\textwidth]{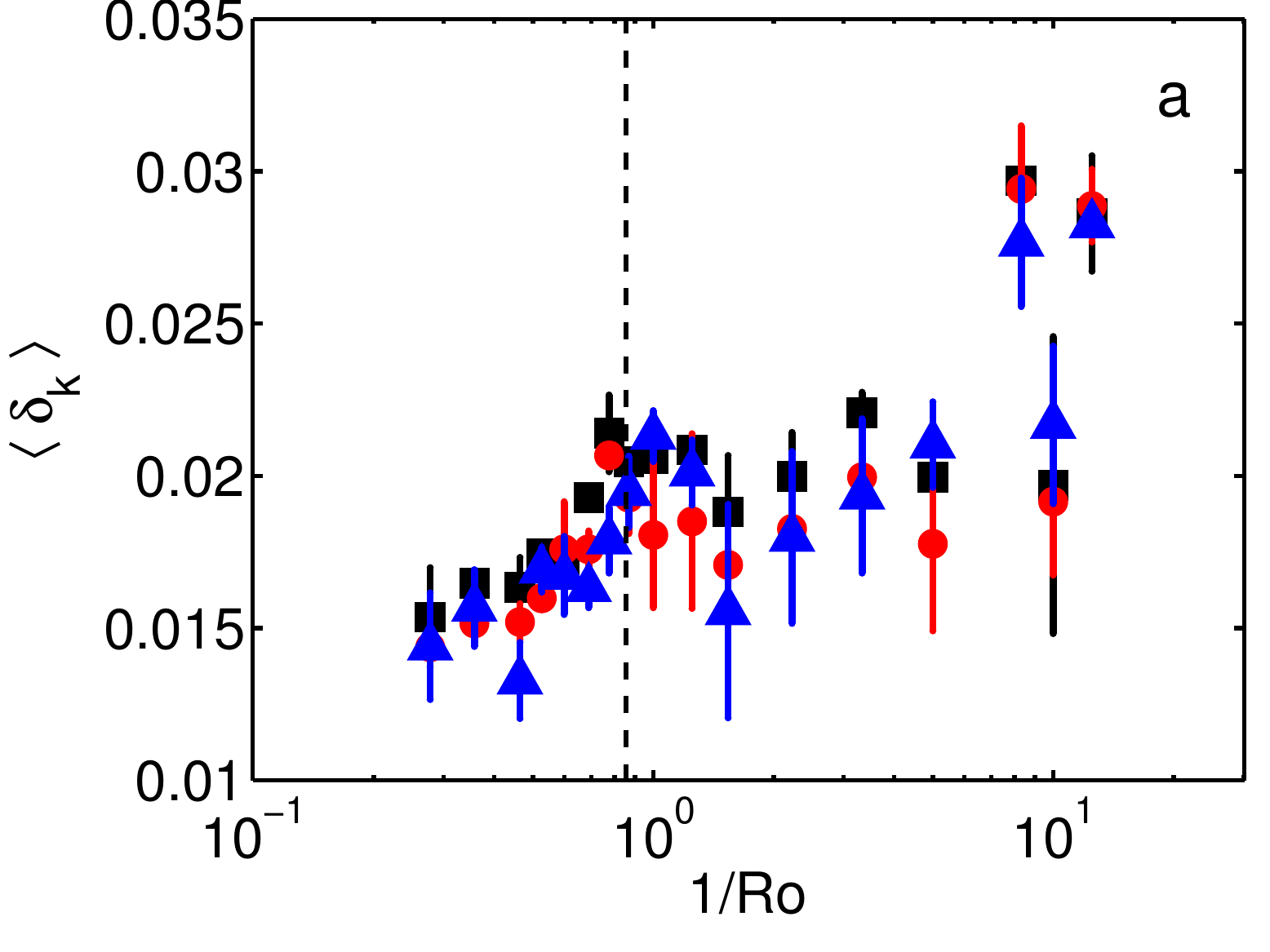}}
  \caption{(color online) The time-averaged temperature amplitudes $\langle \delta_k \rangle$ for all three 
levels $k\in\{b,m,t\}$, 
with the symbols representing $b$ (red circles), $m$ (black squares) and $t$ 
(blue triangles)\} as function of 1/Ro for 
$\Gamma=1/2$ and Ra=$4.52\times10^9$. 
Vertical dashed line as in figure \ref{figure4}. }
  \label{figure5}
\end{figure}
In the simulations we placed $64$ numerical probes at $r=0.95D/2$, which is outside 
the sidewall boundary layer. 
The probes are placed some distance 
from the wall in order to collect data on the velocity components as well, which are 
all zero at the wall. These velocity data are complementary to the temperature 
measurements as the region in which a relatively high (low) temperature is 
measured should correspond to the region with positive (negative) vertical 
velocity. As we do 
not want to see the effect of very small plume events we apply a moving averaging 
filter of $8$ dimensionless time units, see Stevens {\it et al.}\ \cite{ste10c} for details, to the 
temperature measurements of the probes, before we determined $\bar{S}_k$, with 
$k\in\{b,m,t\}$. Figure \ref{figure4}a-c shows that the measured $\bar{S}_k$ at 
$z/L=0.25$, $z/L=0.50$, and $z/L=0.75$ in the simulations agree well with the 
experimental measurements~\cite{wei11b}, which are based on the temperature 
measurements of just $8$ probes embedded in the sidewall instead of the $64$ 
numerical probes in our simulations. The 
error bars in the figure indicate the difference between ${\bar{S}_k}$ obtained using 
the complete time interval with ${\bar{S}_k}$ based on 
the last half of the simulation. Experimental and numerical data obtained in a $\Gamma=1$ 
sample revealed that ${\bar{S}_k}$ strongly decreases to values around ${\bar{S}_k}\approx 0.2$ 
when the heat transport 
enhancement sets in \cite{wei11b}. This is obviously not the case in the $\Gamma=1/2$ 
sample (in general ${\bar{S}_k}\approx 0.5$ for 1 $\lesssim$ 1/Ro $\lesssim$ 10, see 
figure \ref{figure4}). The experimental results do not suggest that this general 
feature depends on Ra~\cite{wei11b}.

Following Ref.~\cite{bro06} the orientation and strength of the LSC can be 
determined by fitting the function $\theta_i = \theta_k + \delta_k \cos(\phi_i-\phi_k)$ to 
the temperatures recorded by the numerical probes at the height $k\in  \{b,m,t \}$, with 
$b$, $m$, and $t$ defined below Eq.~(\ref{Eq Relative Strength LSC}). 
Here, $\phi_i=2 i\pi/N$, with $N$ the number of probes, 
refers to the azimuthal position of the probes, and $\delta_k$ and $\phi_k$ indicate 
the temperature amplitude and orientation of the LSC, respectively. In figure 
\ref{figure5} we plot the time-averaged temperature amplitude $\langle \delta_k\rangle$ of the LSC as 
function of the rotation 
rate for $\Gamma=1/2$. A comparison with the experimental data of Weiss and 
Ahlers \cite{wei11b} (where results are available for both $\Gamma=1/2$ and $\Gamma=1$) 
shows that within our statistical convergence the trends shown in 
the numerical data for $\Gamma=1/2$ are similar to those revealed in the experiments 
(and show a remarkably different trend compared to results for $\Gamma=1$ which show a 
decreasing $\langle \delta_k\rangle$ for increasing rotation~\cite{wei11b}). The temperature 
amplitude of the LSC is relatively small for weak rotation (small values of 1/Ro). It 
becomes slightly larger when increasing the rotation rate from zero to the critical 
rotation rate (1/Ro$_c$ $\approx$ 0.86). Subsequently, a small dip 
just after the onset of heat transport enhancement is observed, which is followed by a 
small further increase in the temperature amplitude.

In summary, in a $\Gamma=1$ sample the transition from the regime 
with no or weak rotation, with the LSC as the dominant flow structure, to the 
rotation dominated regime, with vertically-aligned vortices, is indicated 
by a strong reduction of ${\bar{S}_k}$. Also $\langle \delta_k\rangle$ decreases 
with increasing rotation rate also suggesting disappearance of the LSC. In contrast, 
in a $\Gamma=1/2$ sample these criteria do not provide evidence that the LSC is 
destroyed at the onset of heat 
transport enhancement. In the following section we use data that are only available 
in the simulations to show that also in a $\Gamma=1/2$ sample the LSC is destroyed at 
the onset of heat transport enhancement and that the vortices arrange such that on 
average a sinusoidal temperature profile close to the sidewall is still measured.

\begin{figure}
  \centering
  \subfigure{\includegraphics[width=0.04\textwidth]{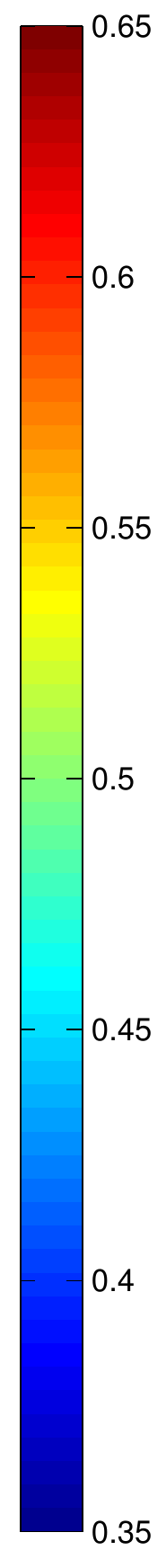}}
  \subfigure{\includegraphics[width=0.18\textwidth]{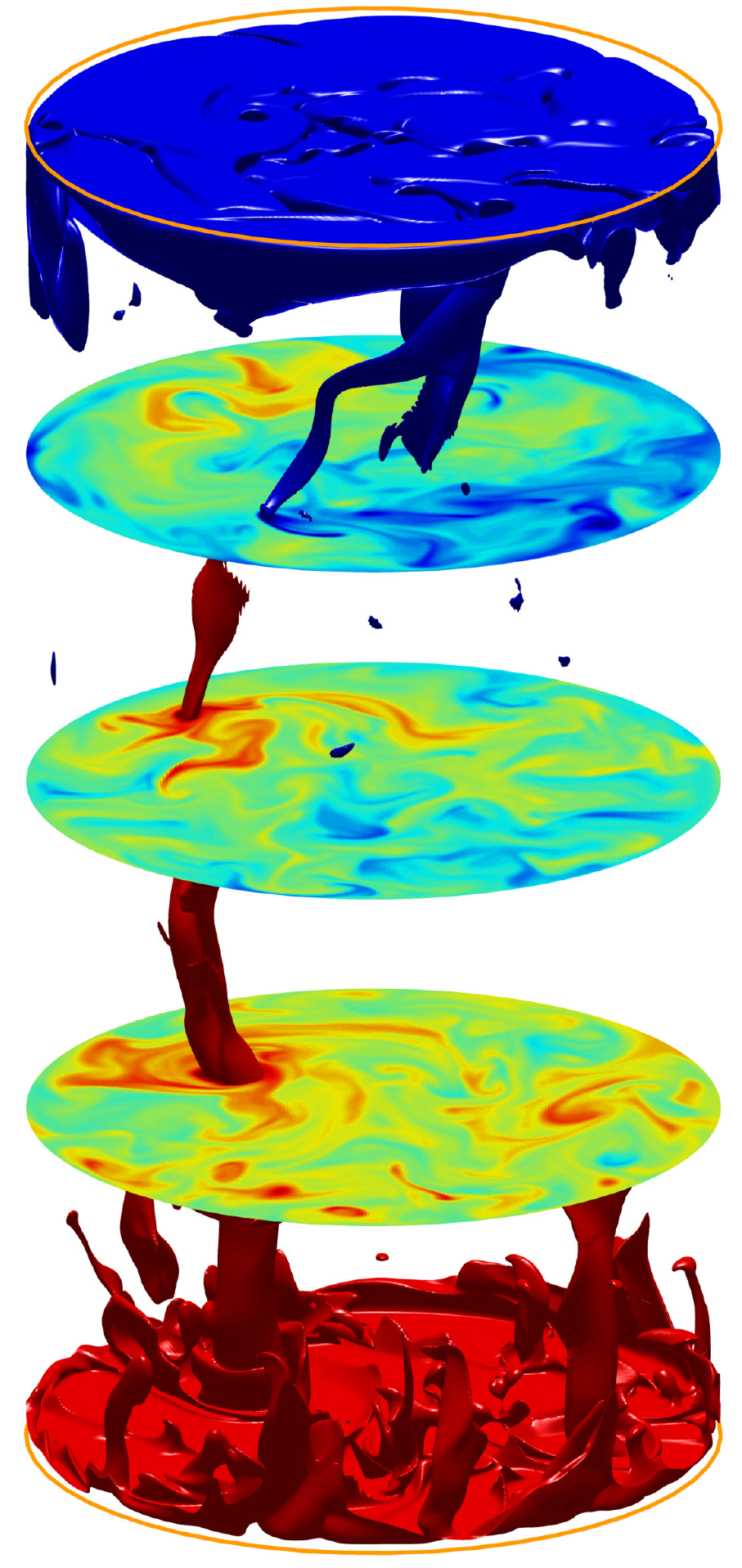}}
  \subfigure{\includegraphics[width=0.18\textwidth]{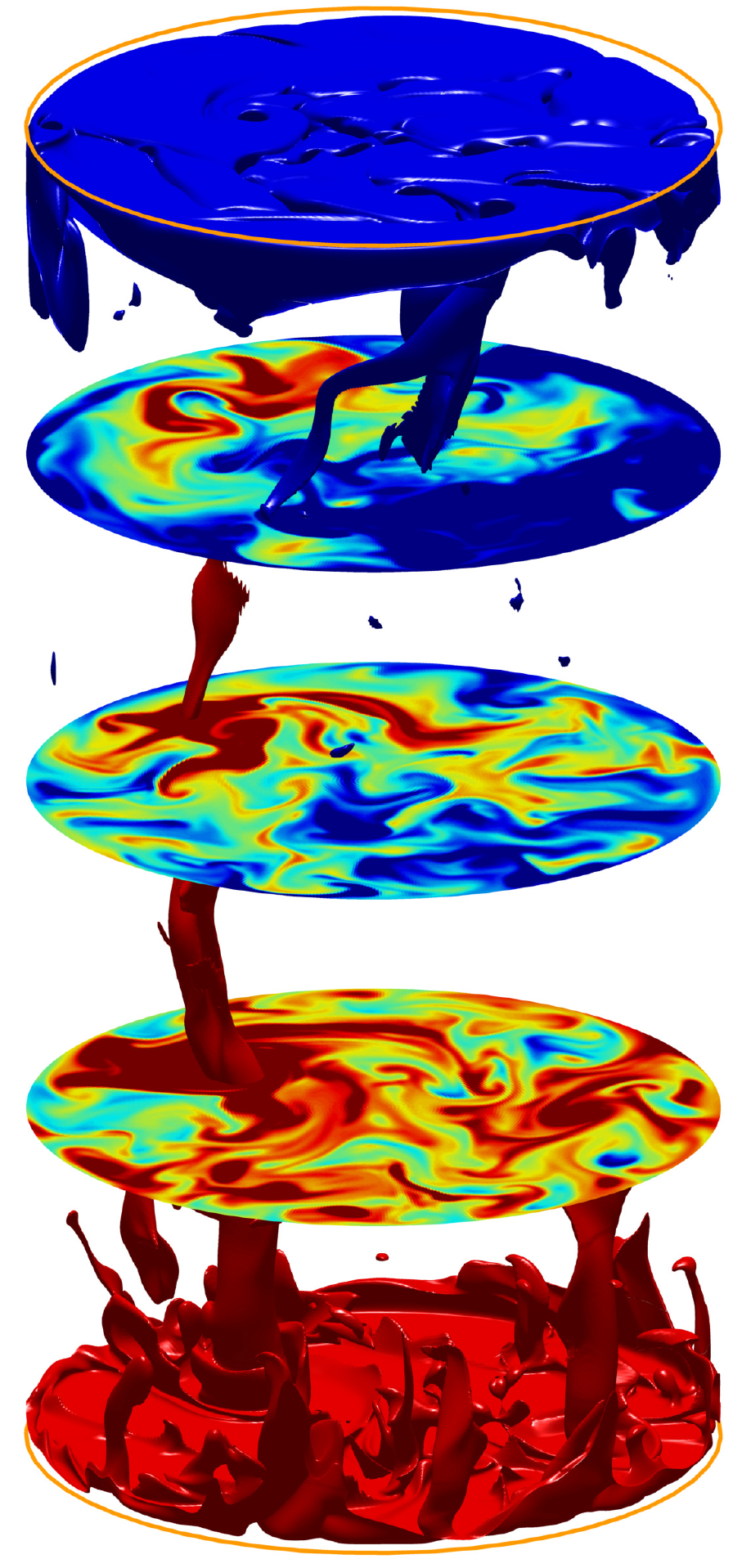}}
  \subfigure{\includegraphics[width=0.04\textwidth]{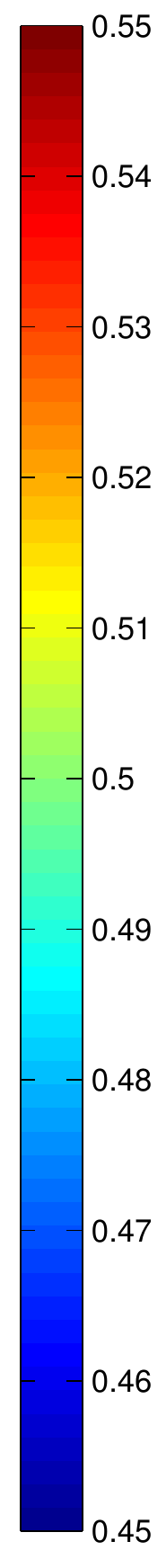}}
  \caption{(color online) Flow visualization for Ra=$4.52\times10^9$, Pr=4.38, and 1/Ro=3.33 in an 
aspect ratio $\Gamma=1/2$ sample. The colormap for the horizontal planes at 
$z/L=0.25$, $z/L=0.50$, and $z/L=0.75$ is different in panel a and b, and is indicated 
on the left and right hand side, respectively. The red temperature isosurface, originating from the bottom,  
indicates the region $\Delta \gtrsim 0.65$, whereas the blue isosurface, originating from the top,  indicates the 
region $\Delta \lesssim 0.35$. Corresponding movies can be found in the supplementary 
material \cite{supplementary}. }
  \label{figure6}
\end{figure}

\begin{figure}
  \centering
  \subfigure{\includegraphics[width=0.22\textwidth]{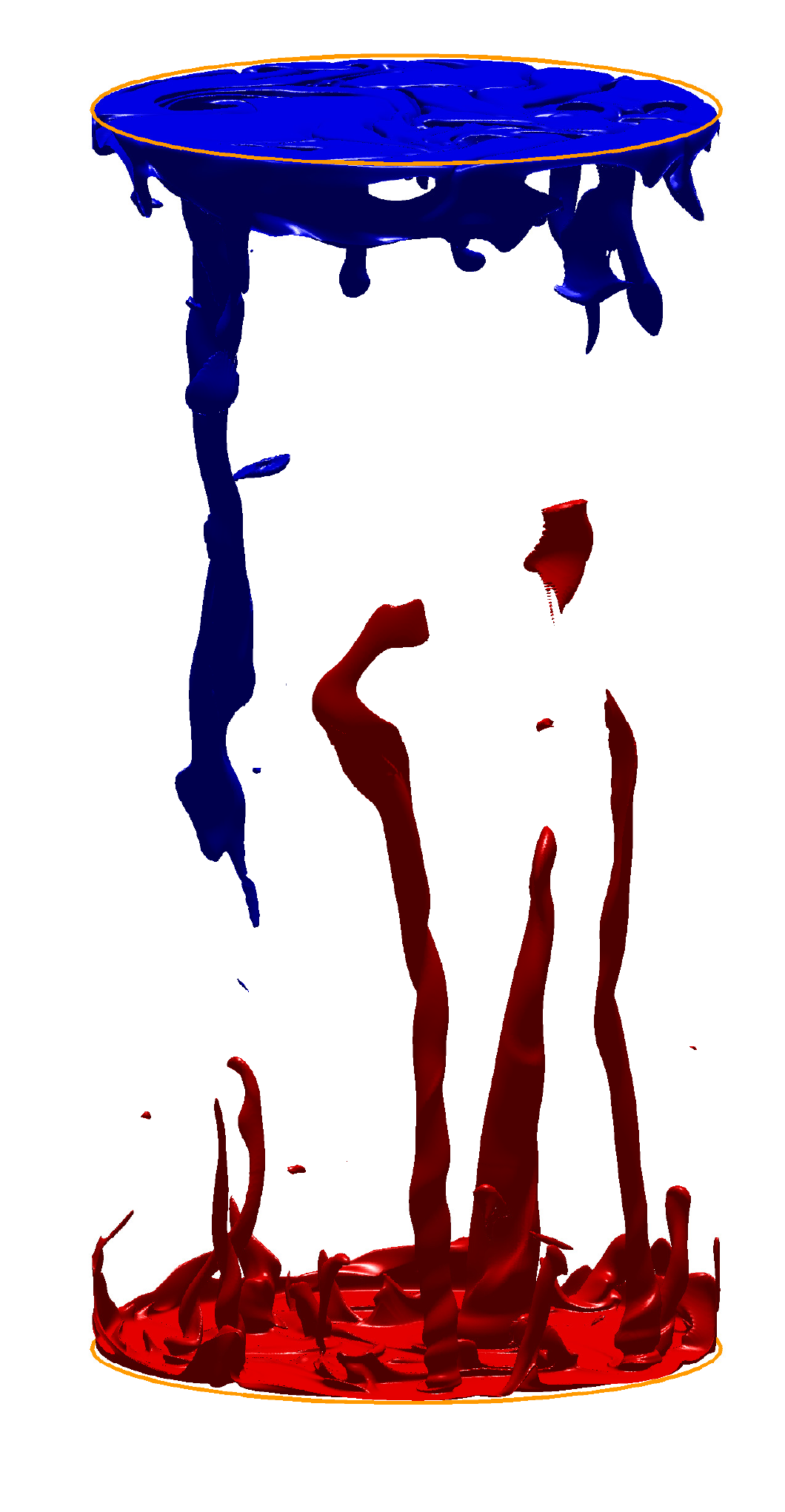}}
  \subfigure{\includegraphics[width=0.22\textwidth]{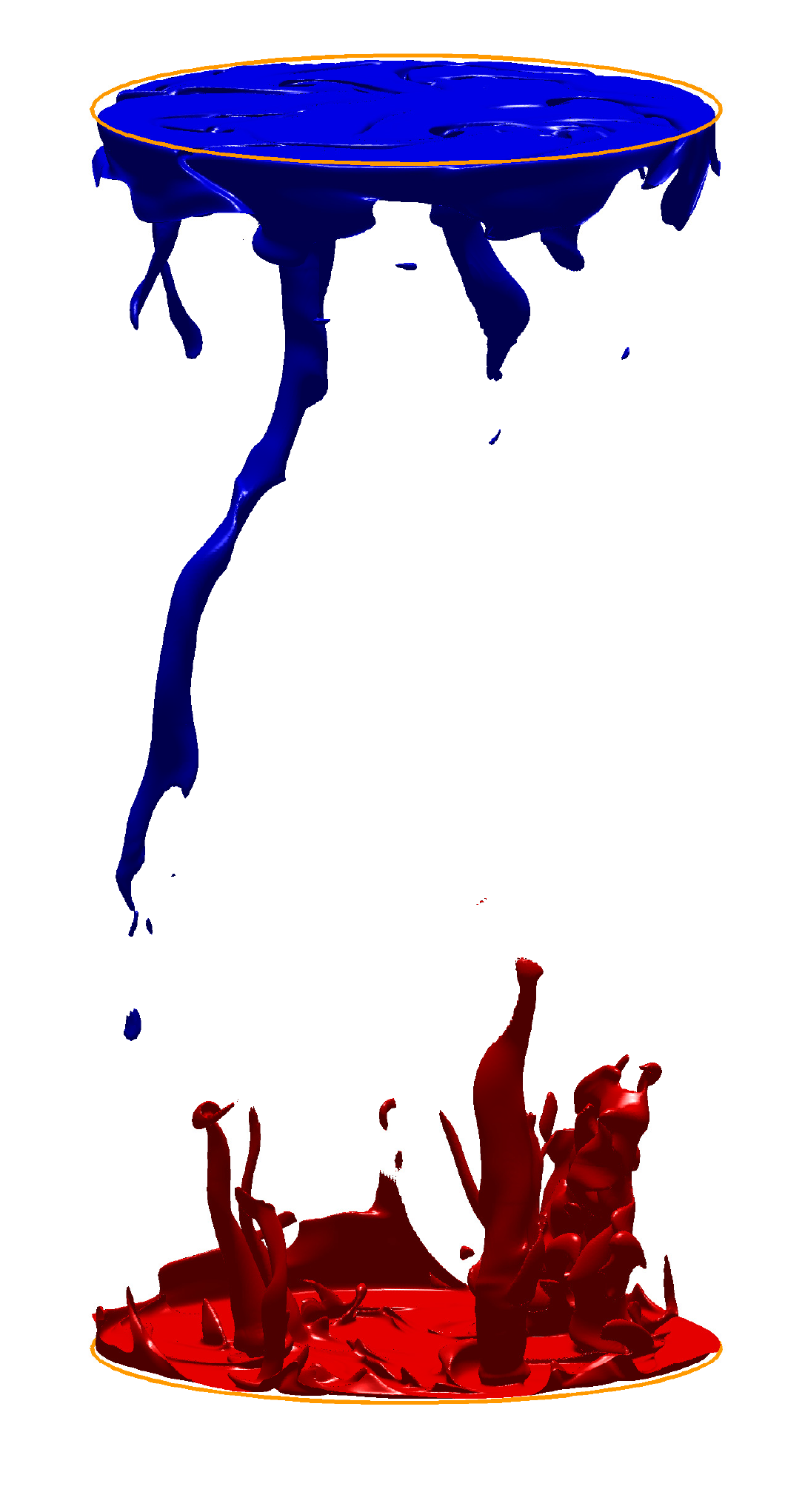}}\\
  \subfigure{\includegraphics[width=0.22\textwidth]{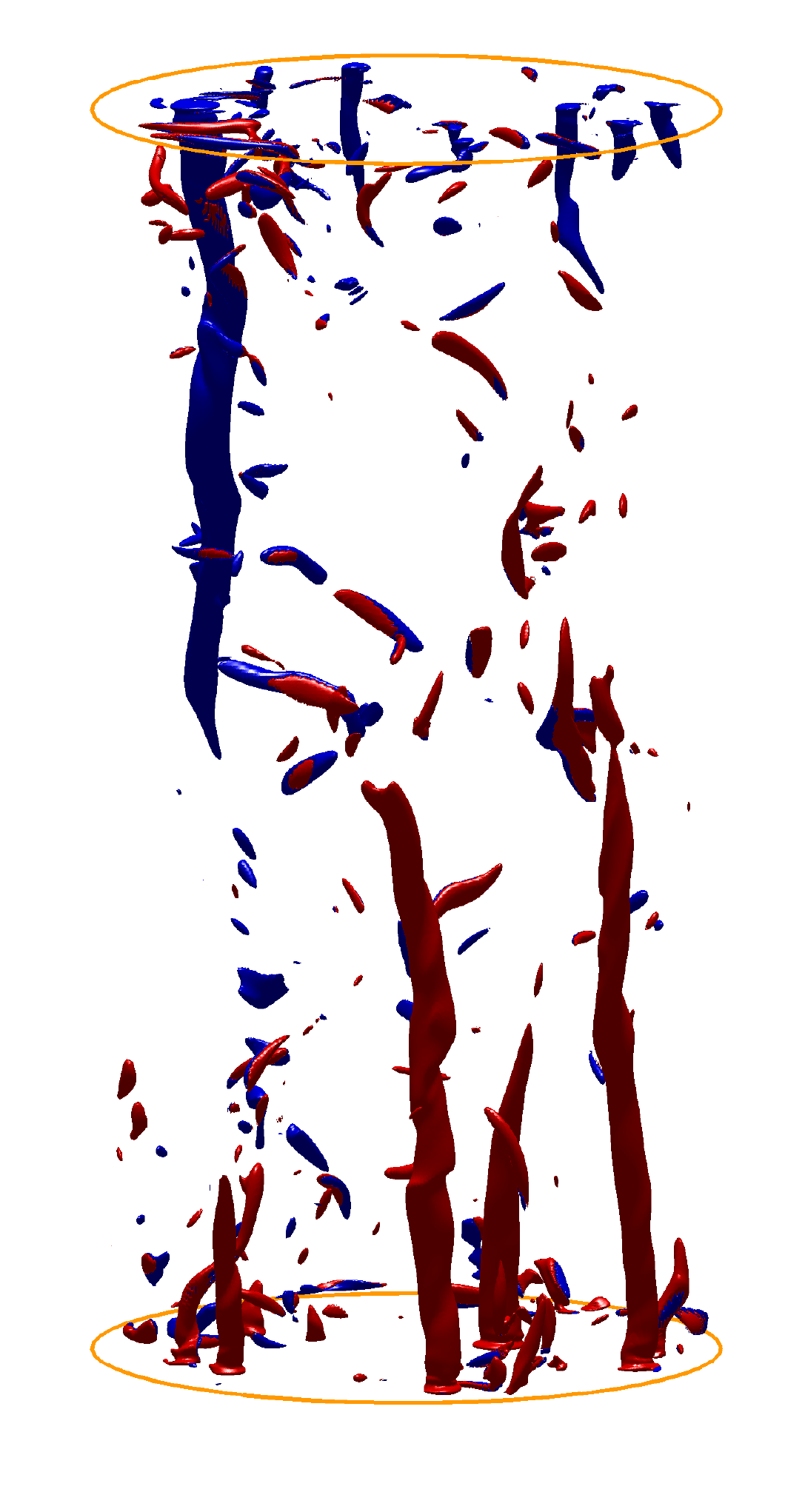}}
  \subfigure{\includegraphics[width=0.22\textwidth]{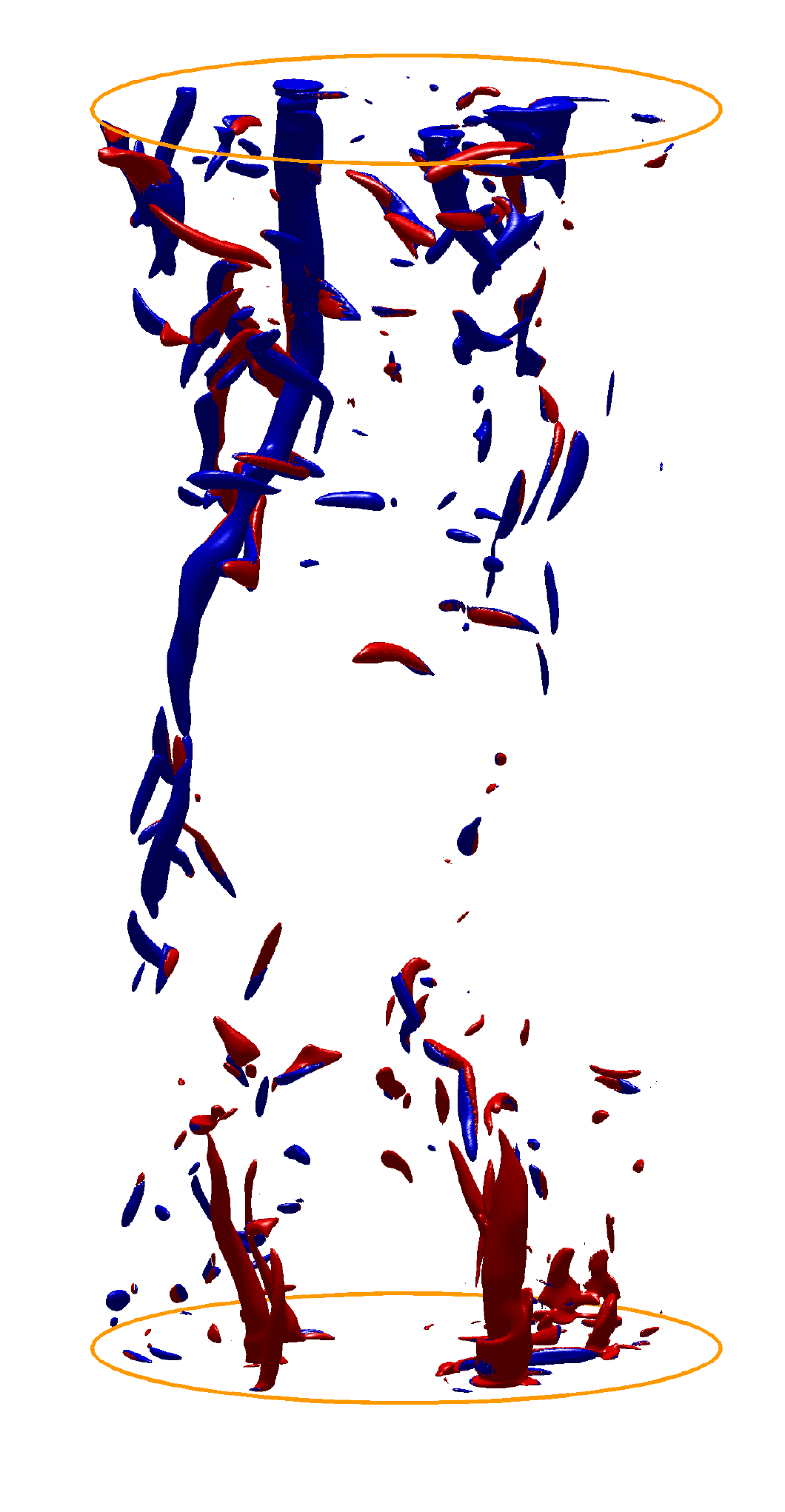}}
  \caption{(color online) Flow visualization for Ra=$4.52\times10^9$, Pr=4.38, and 1/Ro=3.33 in 
an aspect ratio $\Gamma=1/2$ sample The top row shows three-dimensional temperature 
isosurfaces for two different time instances where the red region indicates 
$\Delta \gtrsim 0.65$, whereas the blue isosurfaces indicate the region 
$\Delta \lesssim 0.35$, see figure \ref{figure6}. The bottom row shows a visualization of the 
vortices, based on the $Q_{3D}$ criterion. Note that the red (blue) regions 
indicated by the temperature isosurface indeed correspond to vortex regions.}
  \label{figure7}
\end{figure}

\section{Flow structures after transition towards rotation-dominated regime} \label{chapter4}
In this Section we use the availability of all flow data from the simulations to explain 
why in the rotating regime in an aspect ratio $\Gamma=1/2$ sample the value of ${\bar{S}_k}$ 
does not decrease to values as small as 0.2 or less as usually observed in the 
$\Gamma=1$ case. In particular, we find that the minimum value is ${\bar{S}_k}\approx 0.5$ for 
1 $\lesssim$ 1/Ro $\lesssim$ 10 (see figure \ref{figure4}). Only for 1/Ro $\gtrsim$ 10 the 
experimental data indicate a further substantial decrease of ${\bar{S}_k}$~\cite{wei11b}. Weiss and 
Ahlers already proposed the idea of a two-vortex state which may explain the temperature 
profiles measured in their experiments. However, no experimental validation was possible 
in their experiments. To address this issue we have visualized the flow and temperature 
field for Ra = $4.52 \times 10^9$ and Pr=4.38 at 1/Ro = 3.33, see figures 
\ref{figure6} (isosurfaces with constant temperature) and \ref{figure7} (visualization of 
vortices by means of temperature and vorticity) 
and the complementary movies. The visualization based on temperature isosurfaces reveals 
that also at this high Ra number vertically-aligned vortices are formed in the 
rotation dominated regime. 
In the corresponding movies one can see that close to the 
bottom (top) plate several vortices containing warm (cold) fluid are formed.
Although several vortices 
are created close to the horizontal plates only a few vortices are strong enough to 
reach the horizontal planes (at $z/L=0.25$, $z/L=0.50$ and $z/L=0.75$) so that they can 
be observed by sidewall temperature probes. Therefore, most of the time we see only a few 
prevailing vortices at the horizontal plane at midheight. In addition, we notice the 
non-trivial result that the vortices arrange such that all warm rising vortices tend to 
cluster on one side of the cell and the sinking cold vortices on the opposite side 
(contrary to what is observed for $\Gamma=1$). Note that the 
horizontal planes (at $z/L=0.25$, $z/L=0.50$ and $z/L=0.75$) reveal that the main 
rising and sinking vortices are surrounded by a warm (cold) region. As these warm 
and cold regions are formed on opposite sides of the cell this explains the 
cosine temperature profile that is observed close to the sidewall.\\
To further confirm that indeed vertically-aligned vortices are formed we analyzed the 
flow structure in more detail by determining the positions of the vortices by applying 
a vortex detection algorithm. The fully 
three-dimensional detection technique is based on the velocity gradient tensor 
${\bf \nabla u} = \partial_i u_j (i,j \in {1,2,3})$. This tensor can 
be split into a symmetric and antisymmetric part 
${\bf \nabla u} = \frac{1}{2} [ {\bf \nabla u} + ({\bf \nabla u})^T] + 
\frac{1}{2} [{\bf \nabla u} - ({\bf \nabla u})^T]= S + \Omega$. The $Q_{3D}$ criterion, 
according to Hunt {\it et al.}\ \cite{hun88}, defines a region as vortex when 
$Q_{3D} \equiv \frac{1}{2} (\| \Omega\|^2 - \|S\|^2 ) > 0$, where 
$\| A \| = \sqrt{Tr (AA^T)}$ represents the Euclidean norm of the tensor $A$. 
In practice the threshold value of zero to distinguish between vorticity and 
strain dominated regions in the flow, and thus identifying vortices, is found to 
be unsuitable \cite{kun10b}. Here, we apply a relatively high positive threshold 
value, identical for all presented snapshots, that is 'matched' to the temperature 
isosurfaces that are presented in figure \ref{figure7} (top row). 
The position and size of the detected vortex tubes also agree very well 
with the coherent structures indicated by the temperature isosurfaces, see figure 
\ref{figure7}. This confirms that the warm and cold regions shown by 
temperature isosurfaces are indeed vortices. The main difference between the vortices 
identified with the vortex detection algorithms and the areas indicated by the 
temperature isosurfaces are the smaller vortices in the middle. These small vortices do 
not show up when the temperature criterion is used, since their base is not close to the 
bottom (top) plate where warm (cold) fluid enters the vortices at their base.\\ 
We computed the boundary-layer thickness as function of rotation rate and determined 
averaged root-mean-square velocity fluctuations, as has been done previously for the 
$\Gamma=1$ case~\cite{ste09,ste10b}. Our aim is to confirm directly from flow field data that 
the flow structure in a $\Gamma=1/2$ sample indeed changes from a regime dominated by the 
LSC (1/Ro $\lesssim$ 1/Ro$_c$) to one dominated by vertically-aligned vortices 
(1/Ro $\gtrsim$ 1/Ro$_c$).\\
First of all we determined the kinetic boundary-layer thickness. Like 
in Ref.~\cite{ste10b} for the $\Gamma=1$ sample the kinetic boundary-layer thickness is 
approximately constant before the onset of heat transport enhancement sets in at 
the critical rotation rate 1/Ro$_c$. After the onset the kinetic boundary-layer thickness 
scales as (1/Ro)$^{-0.50}$, which is in agreement with the scaling expected from Ekman 
boundary layer theory. The current data for the $\Gamma=1/2$ (not shown here; graph 
similar as for $\Gamma=1$~\cite{ste09}) thus reveal that the 
boundary-layer structure changes from a Prandtl-Blasius type boundary layer when no or 
weak rotation is applied to an Ekman type boundary layer when 1/Ro$>$1/Ro$_c$. Subsequently, 
we determined the volume-averaged vertical velocity fluctuations, and the vertical 
velocity fluctuations at the edge of the thermal boundary layer. Although not shown here, 
our computations for $\Gamma=1/2$ indicate that the 
volume-averaged vertical velocity fluctuations slightly increase for 1/Ro $\lesssim$ 1/Ro$_c$ 
and strongly decrease for 1/Ro $\gtrsim$ 1/Ro$_c$, thus indicating that the LSC is destroyed. 

The strong decrease in the 
normalized volume-averaged vertical velocity fluctuations coincides with a further 
increase of the horizontal average at the edge of the thermal boundary layers. The 
increase of the fluctuations at the boundary-layer height signifies enhanced Ekman 
transport. Thus these averages provide additional support that the dominant flow 
structure changes from a LSC to a regime dominated by vertically-aligned vortices. 
The average arrangement of these few vortices supports the presence of a sinusoidal 
azimuthal temperature profile at (or close to) the sidewall for $\Gamma=1/2$. This is 
also the essential and non-trivial difference with the $\Gamma=1$ sample, where the 
vertically-aligned vortices are distributed randomly.

\section{Conclusion}
We have shown that for $\Gamma = 1/2$  and 1/Ro $\gtrsim$ 0.86 there is no single-roll 
LSC. Instead, we found a complex state of vortices 
which, in the time average, yields a sinusoidal azimuthal temperature variation 
observed in the experiment over a wide range of rotation rates. We have compared 
data from direct 
numerical simulations performed at Ra=$4.52 \times 10^9$ with Pr=4.38 (water) in an 
aspect ratio $\Gamma=1/2$ sample at different rotation rates with the experimental 
results of Weiss and Ahlers \cite{wei11b}. We find very close agreement in global 
properties, i.e. the measured Nusselt number, and local measurements, i.e. the vertical 
temperature gradient at the sidewall and the behavior of ${\bar{S}_k(1/Ro)}$. In 
contrast to the $\Gamma=1$ case both ${\bar{S}_k}$ and the temperature amplitude 
of the LSC do not indicate any significant changes at the moment that heat transport 
enhancement sets in. Weiss and Ahlers \cite{wei11b} already discussed that ${\bar{S}_k}$ 
cannot distinguish between a single roll state and a two-vortex state, in which one 
vortex extends vertically from the top into the sample interior and brings down cold 
fluid, while another emanates from the bottom and introduces warm fluid.
Here, we resolved this issue and show with a visualization of temperature isosurfaces 
and with the use of advanced vortex detection algorithms that at high Ra the 
formation of a flow structure with basically a few dominant vortices leads to a sinusoidal-like 
azimuthal temperature profile close to the sidewall. This state gives a high value of 
$\bar{S}_k$ at the three measurement heights due to the warm (cold) fluid 
that spreads out in the horizontal direction (due to horizontal transport of 
heat, see Stevens {\it et al.}\ \cite{ste10b}). This smoothens the temperature peak in the azimuthal 
temperature profile at the sidewall. The observation that ${\bar{S}_k}\approx 0.5$ for very large 
rotation rates (up to 1/Ro$\approx$ 10) in a $\Gamma=1/2$ sample indicates that the hot 
and cold vortices {\it on average} must align themselves such that the upgoing (warm) and 
downgoing (cold) vortices are on opposite sides of the cell, but of course a different 
organization can be formed at certain time instances. This organization of the vertically-aligned 
vortices in the $\Gamma=1/2$ sample is a non-trivial difference with the 
$\Gamma=1$ case. There, the vertically-aligned vortices are distributed 
randomly. We are not sure what physical mechanism causes this difference between a $\Gamma=1$ and $\Gamma=1/2$ sample. At the moment we are considering stereoscopic particle image velocimetry measurements in Eindhoven in order to directly visualize the two-vortex state in a $\Gamma=1/2$ RB sample and to obtain statistics over a much longer time domain than in the simulations. We note that the visualization was not possible in the earlier experiments of Weiss and Ahlers \cite{wei11b} as these experiments focused on getting an accurate measurement of the heat transport and hope that these results can answer this question.

In experiments Weiss and Ahlers \cite{wei11b} showed no significant difference between the $\langle E_1 \rangle / \langle E_{tot} \rangle$ as function of 1/Ro (see figure 5 of their paper) for different Ra. Based on this observation we do not expect a strong Ra number dependence of the $\bar{S}(1/Ro)$ curve at a given aspect ratio. At the moment there are no measurements or simulations available that studied the Pr number dependence of $\bar{S}(1/Ro)$ and new measurements of numerical simulations would be necessary to determine whether there is a strong Pr number effect.

From previous investigations on rotating RB convection in $\Gamma=1$ cells it was concluded 
that ${\bar{S}_k}$ is a good indicator for the presence of the LSC. From the results 
reported here we conclude that this quantity cannot provide a unique answer whether the 
LSC is present or not in turbulent rotating convection in $\Gamma=1/2$ samples. The reason for this is that the vortices in a $\Gamma=1/2$ align in such a way that on average a cosine like temperature profile is formed in the azimuthal direction along the sidewall.

\begin{acknowledgments}
\noindent
{\it Acknowledgement:} We benefitted form numerous stimulating discussions with Guenter 
Ahlers and Stephan Weiss and we thank them for providing the (unpublished) data presented 
in figure \ref{figure3} and \ref{figure4}. We thank the DEISA Consortium (www.deisa.eu), 
co-funded through the EU FP7 project RI-222919, for support within the DEISA Extreme 
Computing Initiative. We thank Wim Rijks (SARA) and Siew Hoon Leong (Cerlane) (LRZ) for 
support during the DEISA project. The simulations in this project were performed on the 
Huygens cluster (SARA) and HLRB-II cluster (LRZ). RJAMS was financially supported by the 
Foundation for Fundamental Research on Matter (FOM), which is part of NWO.
\end{acknowledgments}


\begin{thebibliography}{10}

\bibitem{ahl09}
G. Ahlers, S. Grossmann, and D. Lohse, {\em Heat transfer and large scale
  dynamics in turbulent {{Rayleigh-B\'enard}} convection}, Rev. Mod. Phys. {\bf
  81},  503  (2009).

\bibitem{loh10}
D. Lohse and K.~Q. Xia, {\em Small-Scale Properties of Turbulent
  {{Rayleigh-B\'enard}} Convection}, Annu. Rev. Fluid Mech. {\bf 42},  335
  (2010).

\bibitem{bro05b}
E. Brown, A. Nikolaenko, and G. Ahlers, {\em Reorientation of the large-scale
  circulation in turbulent {{Rayleigh-B\'enard}} convection}, Phys. Rev. Lett.
  {\bf 95},  084503  (2005).

\bibitem{xi09}
H.~D. Xi, S.~Q. Zhou, Q. Zhou, T.~S. Chan, and K.~Q. Xia, {\em Origin of
  temperature oscillations in turbulent thermal convection}, Phys. Rev. Lett.
  {\bf 102},  044503  (2009).

\bibitem{ste09}
R.~J. A.~M. Stevens, J.-Q. Zhong, H.~J.~H. Clercx, G. Ahlers, and D. Lohse,
  {\em Transitions between turbulent states in rotating {{Rayleigh-B\'enard}}
  convection}, Phys. Rev. Lett. {\bf 103},  024503  (2009).

\bibitem{ste10c}
R.~J. A.~M. Stevens, H.~J.~H. Clercx, and D. Lohse, {\em Effect of plumes on
  measuring the large-scale circulation in turbulent {{Rayleigh-B\'enard}}
  convection}, Phys. Fluids {\bf 23},  095110  (2011).

\bibitem{ros69}
H.~T. Rossby, {\em A study of {{B\'enard}} convection with and without
  rotation}, J. Fluid Mech. {\bf 36},  309  (1969).

\bibitem{cha81}
S. Chandrasekhar, {\em Hydrodynamic and Hydromagnetic Stability} (Dover, New
  York, 1981).

\bibitem{jul96b}
K. Julien, S. Legg, J. McWilliams, and J. Werne, {\em Hard turbulence in
  rotating {R}ayleigh--{B}{\'e}nard convection}, Phys. Rev. E {\bf 53},  R5557
  (1996).

\bibitem{vor02}
P. Vorobieff and R.~E. Ecke, {\em Turbulent rotating convection: an
  experimental study}, J. Fluid Mech. {\bf 458},  191  (2002).

\bibitem{kun08b}
R.~P.~J. Kunnen, H.~J.~H. Clercx, and B.~J. Geurts, {\em Breakdown of
  large-scale circulation in turbulent rotating convection}, Europhys. Lett.
  {\bf 84},  24001  (2008).

\bibitem{kin09}
E.~M. King, S. Stellmach, J. Noir, U. Hansen, and J.~M. Aurnou, {\em Boundary
  layer control of rotating convection systems}, Nature {\bf 457},  301
  (2009).

\bibitem{zho09b}
J.-Q. Zhong, R.~J. A.~M. Stevens, H.~J.~H. Clercx, R. Verzicco, D. Lohse, and
  G. Ahlers, {\em {{Prandtl}}-, {{Rayleigh}}-, and {{Rossby}}-number dependence
  of heat transport in turbulent rotating {{Rayleigh-B\'enard}} convection},
  Phys. Rev. Lett. {\bf 102},  044502  (2009).

\bibitem{bou90}
B.~M. Boubnov and G.~S. Golitsyn, {\em Temperature and velocity field regimes
  of convective motions in a rotating plane fluid layer}, J. Fluid Mech. {\bf
  219},  215  (1990).

\bibitem{zho93}
F. Zhong, R.~E. Ecke, and V. Steinberg, {\em Rotating {{Rayleigh-B\'enard}}
  convection: asymmetrix modes and vortex states}, J. Fluid Mech. {\bf 249},
  135  (1993).

\bibitem{sak97}
S. Sakai, {\em The horizontal scale of rotating convection in the geostrophic
  regime}, J. Fluid Mech. {\bf 333},  85  (1997).

\bibitem{jul96}
K. Julien, S. Legg, J. McWilliams, and J. Werne, {\em Rapidly rotating
  {{Rayleigh-B\'enard}} convection}, J. Fluid Mech. {\bf 322},  243  (1996).

\bibitem{liu97}
Y. Liu and R.~E. Ecke, {\em Heat transport scaling in turbulent
  {{Rayleigh-B\'enard}} convection: effects of rotation and {{Prandtl}}
  number}, Phys. Rev. Lett. {\bf 79},  2257  (1997).

\bibitem{liu09}
Y. Liu and R.~E. Ecke, {\em Heat transport measurements in turbulent rotating
  {{Rayleigh-B\'enard}} convection}, Phys. Rev. E {\bf 80},  036314  (2009).

\bibitem{zho10c}
J.-Q. Zhong and G. Ahlers, {\em Heat transport and the large-scale circulation
  in rotating turbulent {{Rayleigh-B\'enard}} convection}, J. Fluid Mech. {\bf
  665},  300  (2010).

\bibitem{wei10}
S. Weiss, R.~J. A.~M. Stevens, J.-Q. Zhong, H.~J.~H. Clercx, D. Lohse, and G.
  Ahlers, {\em Finite-size effects lead to supercritical bifurcations in
  turbulent rotating {{Rayleigh-B\'enard}} convection}, Phys. Rev. Lett. {\bf
  105},  224501  (2010).

\bibitem{wei11}
S. Weiss and G. Ahlers, {\em Heat transport by turbulent rotating
  {{Rayleigh-B\'enard}} convection and its dependence on the aspect ratio}, J.
  Fluid. Mech. {\bf 684},  407  (2011).

\bibitem{wei11b}
S. Weiss and G. Ahlers, {\em The large-scale flow structure in turbulent
  rotating {{Rayleigh-B\'enard}} convection}, J. Fluid. Mech. {\bf 688},  461
  (2011).

\bibitem{kun11}
R.~P.~J. Kunnen, R.~J. A.~M. Stevens, J. Overkamp, C. Sun, G.~J.~F. van Heijst,
  and H.~J.~H. Clercx, {\em The role of {{Stewartson}} and {{Ekman}} layers in
  turbulent rotating {{Rayleigh-B\'enard}} convection}, J. Fluid. Mech. {\bf
  688},  422  (2011).

\bibitem{kun10}
R.~P.~J. Kunnen, B.~J. Geurts, and H.~J.~H. Clercx, {\em Experimental and
  numerical investigation of turbulent convection in a rotating cylinder}, J.
  Fluid Mech. {\bf 642},  445  (2010).

\bibitem{kun10b}
R.~P.~J. Kunnen, B.~J. Geurts, and H.~J.~H. Clercx, {\em Vortex statistics in
  turbulent rotating convection}, Phys. Rev. E {\bf 82},  036306  (2010).

\bibitem{ste10a}
R.~J. A.~M. Stevens, H.~J.~H. Clercx, and D. Lohse, {\em Optimal {{Prandtl}}
  number for heat transfer in rotating {{Rayleigh-B\'enard}} convection}, New
  J. Phys. {\bf 12},  075005  (2010).

\bibitem{ste10b}
R.~J. A.~M. Stevens, H.~J.~H. Clercx, and D. Lohse, {\em Boundary layers in
  rotating weakly turbulent {{Rayleigh-B\'enard}} convection.}, Phys. Fluids
  {\bf 22},  085103  (2010).

\bibitem{ste11b}
R.~J. A.~M. Stevens, J. Overkamp, D. Lohse, and H.~J.~H. Clercx, {\em Effect of
  aspect-ratio on vortex distribution and heat transfer in rotating
  {{Rayleigh-B\'enard}}}, Phys. Rev. E {\bf 84},  056313  (2011).

\bibitem{liu11}
Y. Liu and R.~E. Ecke, {\em Local temperature measurements in turbulent
  rotating {{Rayleigh-B\'enard}} convection}, Phys. Rev. E {\bf 84},  016311
  (2011).

\bibitem{kin12}
E.~M. King, S. Stellmach, and J.~M. Aurnou, {\em Heat transfer by rapidly
  rotating {{Rayleigh-B\'enard}} convection}, J. Fluid Mech. {\bf 691},  568
  (2012).

\bibitem{sch09}
S. Schmitz and A. Tilgner, {\em Heat transport in rotating convection without
  {{Ekman}} layers}, Phys. Rev. E {\bf 80},  015305  (2009).

\bibitem{sch10}
S. Schmitz and A. Tilgner, {\em Transitions in turbulent rotating
  {{Rayleigh-B\'enard}} convection}, Geophysical and Astrophysical Fluid
  Dynamics {\bf 104},  481  (2010).

\bibitem{har99}
J.~E. Hart and D.~R. Olsen, {\em On the thermal offset in turbulent rotating
  convection}, Phys. Fluids {\bf 11},  2101  (1999).

\bibitem{wei10b}
S. Weiss and G. Ahlers, {\em Turbulent {{Rayleigh-B\'enard}} convection in a
  cylindrical container with aspect ratio {{$\Gamma$=0.50}} and {{Prandtl}}
  number {{Pr = 4.38}}}, J. Fluid. Mech. {\bf 676},  5   (2011).

\bibitem{ver96}
R. Verzicco and P. Orlandi, {\em A finite-difference scheme for
  three-dimensional incompressible flow in cylindrical coordinates}, J. Comput.
  Phys. {\bf 123},  402  (1996).

\bibitem{ver99}
R. Verzicco and R. Camussi, {\em Prandtl number effects in convective
  turbulence}, J. Fluid Mech. {\bf 383},  55  (1999).

\bibitem{ste10}
R.~J. A.~M. Stevens, R. Verzicco, and D. Lohse, {\em Radial boundary layer
  structure and {{Nusselt}} number in {{Rayleigh-B\'enard}} convection}, J.
  Fluid. Mech. {\bf 643},  49å  (2010).

\bibitem{shi10}
O. Shishkina, R.~J. A.~M. Stevens, S. Grossmann, and D. Lohse, {\em Boundary
  layer structure in turbulent thermal convection and its consequences for the
  required numerical resolution}, New J. Phys. {\bf 12},  075022  (2010).

\bibitem{bro06}
E. Brown and G. Ahlers, {\em Rotations and cessations of the large-scale
  circulation in turbulent {{Rayleigh-B\'enard}} convection}, J. Fluid Mech.
  {\bf 568},  351  (2006).

\bibitem{hun88}
J.~C.~R. Hunt, A. Wray, and P. Moin,  {\em Eddies, stream, and convergence zones in turbulent flows}, Report CTR-S88, Center for Turbulence
  Research (unpublished).
  
\bibitem{supplementary}
See Supplemental Material at [URL will be inserted by publisher] for movie showing the evolution of the main flow structures over time.
  

\end{thebibliography}

\end{document}